\documentclass[11pt]{article}
\usepackage[left=1in,right=1in,top=1in]{geometry}

\usepackage[T1]{fontenc}
\usepackage{times}

\usepackage{amsmath,amssymb,amsthm,array,authblk,enumitem,hyperref,multirow,url,xspace}

\usepackage[noend,linesnumbered,ruled,lined,boxed]{algorithm2e}
\SetKwInput{Parameter}{Parameter}
\SetKw{Break}{break}

\usepackage{graphicx}
\DeclareGraphicsExtensions{.eps,.pdf}
\usepackage[caption=false]{subfig}

\usepackage[sort,numbers]{natbib}

\newtheorem{definition}{Definition}
\newtheorem{example}{Example}
\newtheorem{lemma}{Lemma}
\newtheorem{theorem}{Theorem}

\newcommand{\query}{$k$-SIR\xspace}
\newcommand{\rl}{\mathsf{RL}}
\newcommand{\mtts}{MTTS\xspace}
\newcommand{\mttd}{MTTD\xspace}

\DeclareMathOperator*{\argmax}{argmax}

\begin{document}
\title{Semantic and Influence aware k-Representative Queries over Social Streams}

\author[1]{Yanhao Wang}
\author[2]{Yuchen Li}
\author[1]{Kian-Lee Tan}
\affil[1]{School of Computing, National University of Singapore, Singapore}
\affil[2]{School of Information Systems, Singapore Management University, Singapore}
\affil[ ]{$^1$\textit{\{yanhao90, tankl\}@comp.nus.edu.sg\quad $^2$yuchenli@smu.edu.sg}}
\maketitle

\begin{abstract}
Massive volumes of data continuously generated on social platforms
have become an important information source for users.
A primary method to obtain fresh and valuable information
from social streams is \emph{social search}.
Although there have been extensive studies on social search,
existing methods only focus on the \emph{relevance}
of query results but ignore the \emph{representativeness}.
In this paper, we propose a novel Semantic and Influence aware
$k$-Representative ($k$-SIR) query for social streams based on topic modeling.
Specifically, we consider that both user queries and elements are represented
as vectors in the topic space. A $k$-SIR query retrieves
a set of $k$ elements with the maximum \emph{representativeness}
over the sliding window at query time w.r.t. the query vector.
The representativeness of an element set comprises both semantic and influence
scores computed by the topic model.
Subsequently, we design two approximation algorithms, namely
\textsc{Multi-Topic ThresholdStream} (MTTS) and
\textsc{Multi-Topic ThresholdDescend} (MTTD),
to process $k$-SIR queries in real-time.
Both algorithms leverage the ranked lists maintained on each topic
for $k$-SIR processing with theoretical guarantees.
Extensive experiments on real-world datasets demonstrate
the effectiveness of $k$-SIR query compared with existing methods
as well as the efficiency and scalability of our proposed algorithms for $k$-SIR processing.
\end{abstract}

\section{Introduction}
\label{sec:intro}

Enormous amount of data is being continuously generated by web
users on social platforms at an unprecedented rate.
For example, around 650 million tweets are posted by 330 million users on Twitter per day.
Such user generated data can be modeled as continuous social streams,
which are key sources of fresh and valuable information.
Nevertheless, social streams are extremely overwhelming for their huge volumes
and high velocities. It is impractical for users to consume social
data in its raw form.
Therefore, \emph{social
search}~\cite{Chen:2011:TI,Busch:2012:Earlybird,Shraer:2013:TPS,Wu:2013:LSII,Li:2015:Real,U:2017:CTM,Chen:2015:DTP,Li:2016:CAR,Zhang:2017:PLQ}
has become the primary approach to facilitating users on finding their
interested content from massive social streams.

\begin{figure*}
  \centering
  \begin{minipage}{0.98\textwidth}
    \centering
    \small
    \begin{tabular}{|c|c|c|}
    \hline
    \textbf{ID} & \textbf{Tweet} & \textbf{Retweets} \\
    \hline
    $e_{1}$ & @asroma win but it's @LFC joining @realmadrid in the \#UCL final & 3154 \\
    $e_{2}$ & \#OnThisDay in 1993, @ManUtd were crowned the first \#PL champion & 1476 \\
    $e_{3}$ & @Cavs defeats @Raptors 128-110 and leads the series 2-0 in \#NBAPlayoffs & 2706 \\
    $e_{4}$ & LeBron is great! \#NBAPlayoffs & 2 \\
    $e_{5}$ & Congratulations to @LFC reaching \#UCL Final!! \#YNWA & 2167 \\
    $e_{6}$ & LeBron is the 1st player with 40+ points 14+ assists in an \#NBAPlayoffs game & 3489 \\
    $e_{7}$ & Hope this post inspires us to win \#PL champions again in 2018-19 & 4 \\
    $e_{8}$ & Schedule for \#PL and \#NBAPlayoffs tonight & 25 \\
    \hline
    \end{tabular}
    \caption{A list of exemplar tweets}
    \label{fig:motivation:tweets}
    \vspace{1em}
  \end{minipage}%
  \newline
  \begin{minipage}{0.4\textwidth}
    \centering
    \includegraphics[width=0.78\textwidth]{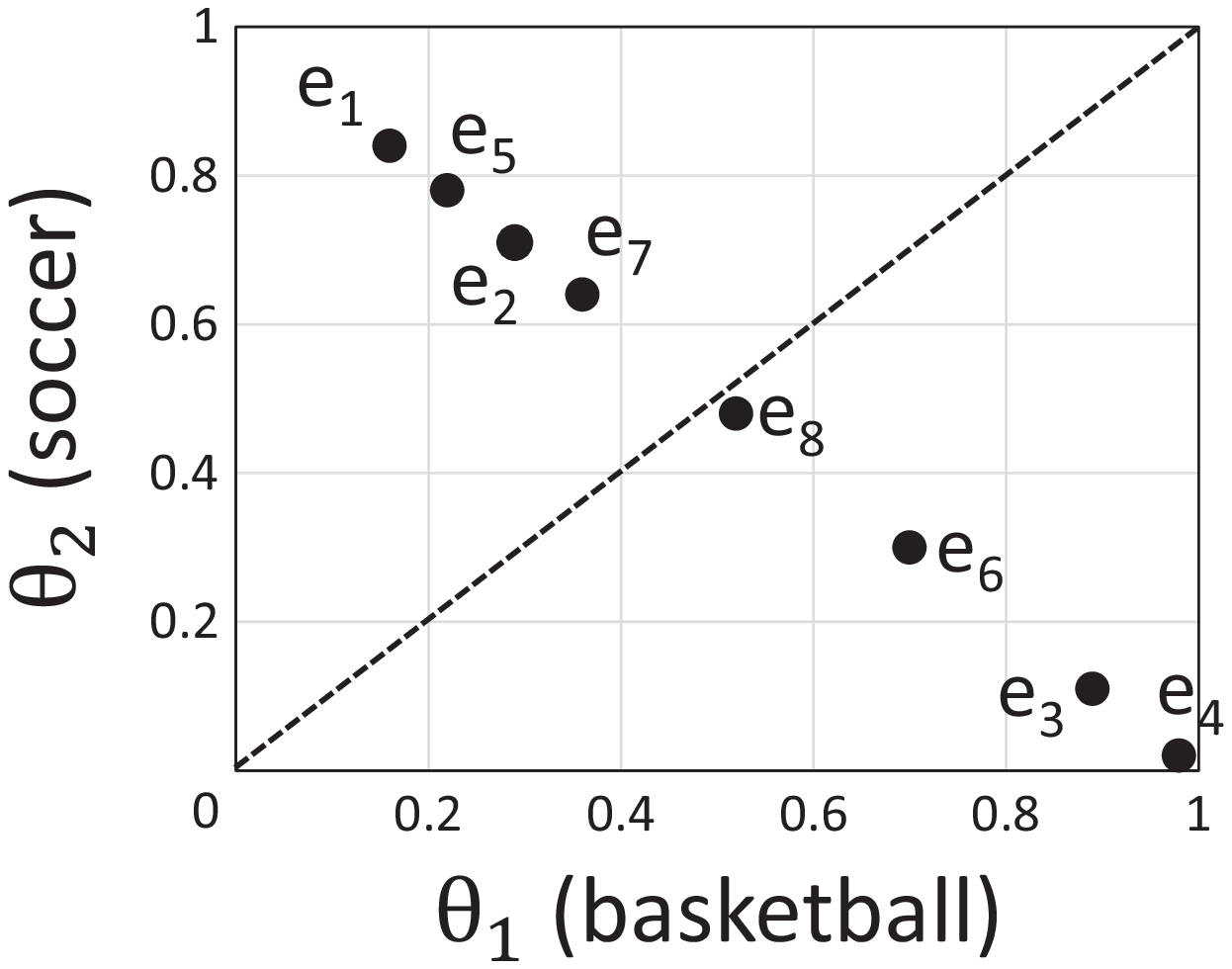}
    \caption{Topic distribution}
    \label{fig:motivation:topics}
  \end{minipage}%
  \begin{minipage}{0.4\textwidth}
    \centering
    \includegraphics[width=\textwidth]{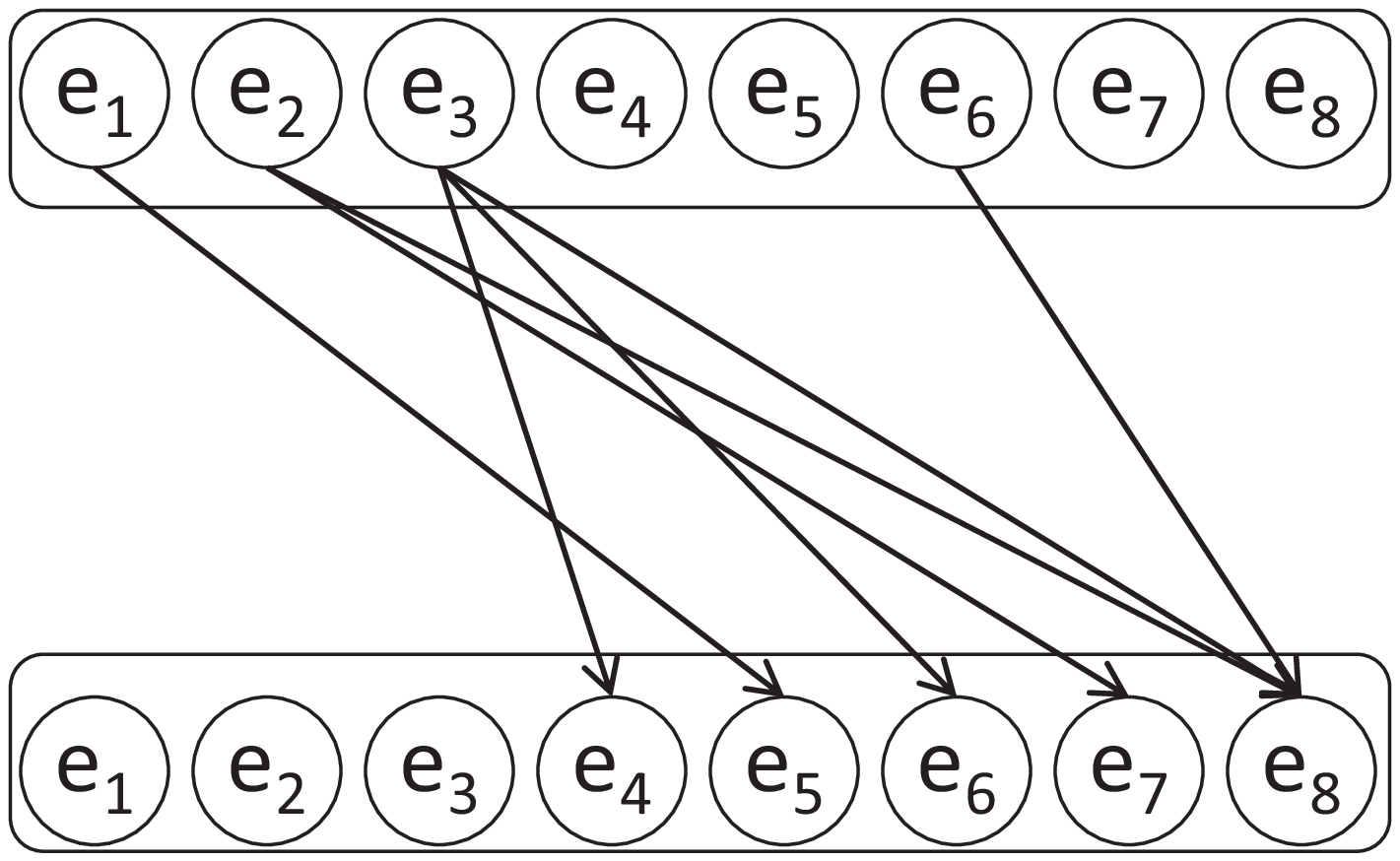}
    \caption{References}
    \label{fig:motivation:refs}
  \end{minipage}
\end{figure*}

Existing search methods for social data can be categorized into keyword-based approaches
and topic-based approaches based on how they measure the relevance between queries and elements.
Keyword-based approaches~\cite{Chen:2011:TI,Busch:2012:Earlybird,Shraer:2013:TPS,Wu:2013:LSII,Li:2015:Real,U:2017:CTM,Chen:2015:DTP}
adopt the \emph{textual relevance} (e.g., TF-IDF and BM25) for evaluation.
However, they merely capture the \emph{syntactic} correlation but ignore the \emph{semantic} correlation.
Considering the tweets in Figure~\ref{fig:motivation:tweets},
if a query ``soccer'' is issued, no results will be found
because none of the tweets contains the term ``soccer''.
It is noted that the words like ``asroma'' and  ``LFC''
are semantically relevant to ``soccer''.
Therefore, elements such as $e_1,e_2$ are relevant
to the query but missing from the result.
Thus, overlooking the semantic meanings of user queries
may degrade the result quality,
especially against social data
where lexical variation is prevalent~\cite{DBLP:conf/icde/HuaWWZZ15}.

To overcome this issue, topic-based approaches~\cite{Li:2016:CAR,Zhang:2017:PLQ}
project user queries and elements into the same latent space
defined by a probabilistic topic model~\cite{Blei:2012:PTM}.
Consequently, queries and elements are both represented as vectors
and their relevance is computed by similarity measures
for vectors (e.g., cosine distance) in the topic space.
Although topic-based approaches can better capture the \emph{semantic} correlation
between queries and elements,
they focus on the \emph{relevance} of results but neglect the \emph{representativeness}.
Typically, they retrieve top-$k$ elements that are the most coherent with the query as the result.
Such results may not be \emph{representative}
in the sense of \emph{information coverage} and \emph{social influence}.
First, users are more satisfied with the results that achieve an extensive coverage of information on query topics
than the ones that provide limited information.
For example, a top-$2$ query on topic $\theta_1$ in Figure~\ref{fig:motivation:topics}
returns $\{e_3,e_4\}$ as the result.
Nevertheless, compared with $e_4$, $e_6$ can provide richer information
to complement the news reported by $e_3$.
Therefore, in addition to \emph{relevance},
it is essential to consider \emph{information coverage} to improve the result quality.
Second, \emph{influence} is another key characteristic to
measure the \emph{representativeness} of social data.
Existing methods for social search~\cite{Wu:2013:LSII,Chen:2011:TI,Li:2016:CAR,Busch:2012:Earlybird}
have taken into account the influences of elements for scoring and ranking.
These methods simply use the influences of authors (e.g., PageRank~\cite{Page:1999:Pagerank} scores)
or the retweet/share count to compute the influence scores.
Such a na{\"i}ve integration of
influence is topic-unaware and may lead to undesired query results.
For example, $e_6$ in Figure~\ref{fig:motivation:tweets},
which is mostly related to $\theta_1$, may appear in
the result for a query on $\theta_2$
because of its high retweet count.
In addition, they do not consider that the influences of elements
evolve over time, when previously trending contents may become outdated
and new posts continuously emerge.
Hence, incorporating
a topic-aware and time-critical influence metric is imperative to capture recently trending elements.

To tackle the problems of existing search methods,
we define a novel Semantic and Influence aware $k$-Representative
(\query) query for social streams based on topic modeling~\cite{Blei:2012:PTM}.
Specifically, a \query query retrieves a set of $k$ elements
from the active elements corresponding to the sliding window $W_t$ at the query time $t$.
The result set collectively
achieves the maximum representativeness score w.r.t. the query vector $\mathbf{x}$,
each dimension of which indicates the degree of interest on a topic.
We advocate the representativeness score of an element set
to be a weighted sum of its semantic and influence
scores on each topic.
We adopt a weighted word coverage model to compute the semantic score
so as to achieve the best information preservation,
where the weight of a word is evaluated based on
its information entropy~\cite{Zhuang:2016:DSS,Nguyen:2017:Retaining}.
The influence score is computed by a probabilistic coverage model
where the influence probabilities are topic-aware.
In addition, we restrict the influences within the sliding window $W_t$
so that the recently trending elements can be selected.

The challenges of real-time \query processing are two-fold.
First, the \query query is \emph{NP-hard}.
Second, it is highly \emph{dynamic},
i.e., the results vary with query vectors and evolve quickly over time.
Due to the submodularity of the scoring function,
existing submodular maximization algorithms,
e.g., CELF~\cite{Leskovec:2007:Cost} and SieveStreaming~\cite{Badanidiyuru:2014:Streaming},
can provide approximation results for \query queries with theoretical guarantees.
However, existing algorithms need to evaluate all active elements at least once for a single query
and often take several seconds to process one \query query as shown in our experiments.
To support real-time \query processing over social streams,
we maintain the ranked lists to sort the active elements on each topic
by topic-wise representativeness score.
We first devise the \textsc{Multi-Topic ThresholdStream} (\mtts) algorithm for \query processing.
Specifically, to prune unnecessary evaluations, \mtts sequentially retrieves elements
from the ranked lists in decreasing order of their scores w.r.t. the query vector
and can be terminated early whenever possible.
Theoretically, it provides $(\frac{1}{2}-\varepsilon)$-approximation results
for \query queries and evaluates each active element at most once.
Furthermore, we propose the \textsc{Multi-Topic ThresholdDescend} (\mttd) algorithm to improve upon \mtts.
\mttd maintains the elements retrieved from ranked lists in a buffer
and permits to evaluate an element more than once to improve the result quality.
Consequently, it achieves a better $(1-\frac{1}{e}-\varepsilon)$-approximation
but has a higher worst-case time complexity than \mtts.
Despite this, \mttd shows better empirical efficiency and result quality than those of \mtts.

Finally, we conduct extensive experiments on three real-world datasets to evaluate
the effectiveness of \query as well as the efficiency and scalability of \mtts and \mttd.
The results of a user study and quantitative analysis demonstrate that
\query achieves significant improvements over existing methods in terms of
\emph{information coverage} and \emph{social influence}.
In addition, \mtts and \mttd achieve up to 124x and 390x speedups over the baselines
for \query processing with at most $5\%$ and $1\%$ losses in quality.

Our contributions in this work are summarized as follows.
\begin{itemize}
  \item We define the \query query to retrieve representative
  elements over social streams where both semantic and influence
  scores are considered. (Section~\ref{sec:problem})
  \item We propose \mtts and \mttd to process \query queries in real-time
  with theoretical guarantees. (Section~\ref{sec:framework})
  \item We conduct extensive experiments to demonstrate the effectiveness
  of \query as well as the efficiency and scalability of our proposed algorithms
  for \query processing. (Section~\ref{sec:experiments})
\end{itemize}

\section{Related Work}
\label{sec:related}

\textbf{Search Methods for Social Streams.}
Many methods have been proposed for searching on social streams.
Here we categorize existing methods into keyword-based approaches and topic-based approaches.

Keyword-based approaches~\cite{Chen:2011:TI,Busch:2012:Earlybird,Shraer:2013:TPS,Wu:2013:LSII,Li:2015:Real,U:2017:CTM,Chen:2015:DTP,Zhang:2017:CIJ}
typically define \emph{top-$k$ queries} to retrieve $k$ elements
with the highest scores as the results where the scoring functions
combine the \emph{relevance} to query keywords (measured by TF-IDF or BM25)
with other contexts such as
\emph{freshness}~\cite{Shraer:2013:TPS,Li:2015:Real,U:2017:CTM,Wu:2013:LSII},
\emph{influence}~\cite{Chen:2011:TI,Wu:2013:LSII},
and \emph{diversity}~\cite{Chen:2015:DTP}.
They also design different indices to support instant updates
and efficient top-$k$ query processing.
However, keyword queries are substantially different from the \query query
and thus keyword-based methods cannot be trivially adapted
to process \query queries based on topic modeling.

As the metrics for textual relevance cannot fully represent
the semantic relevance between user interest and text,
recent work~\cite{Li:2016:CAR,Zhang:2017:PLQ}
introduces topic models~\cite{Blei:2012:PTM} into social search,
where user queries and elements are modeled as vectors in the topic space.
The relevance between a query and an element is measured by cosine similarity.
They define \emph{top-$k$ relevance query}
to retrieve $k$ most relevant elements to a query vector.
However, existing methods typically consider
the \textit{relevance} of results but ignore the \textit{representativeness}.
Therefore, the algorithms in~\cite{Li:2016:CAR,Zhang:2017:PLQ}
cannot be used to process \query queries that emphasize the \textit{representativeness}
of results.

\textbf{Social Stream Summarization.}
There have been extensive studies on social stream
summarization~\cite{DBLP:conf/edbt/AgarwalR17,DBLP:conf/sigir/ShouW0013,DBLP:journals/tkde/WangS00M15,DBLP:conf/eacl/Olariu14,DBLP:journals/tmm/BianYZC15,DBLP:conf/cikm/RenIAR16,DBLP:conf/dasfaa/SongZBS17,DBLP:conf/sigir/RenLMR13}
: the problem of extracting a set of \emph{representative} elements from social streams.
Shou et al.~\cite{DBLP:conf/sigir/ShouW0013,DBLP:journals/tkde/WangS00M15} propose
a framework for social stream summarization based on dynamic clustering.
Ren et al.~\cite{DBLP:conf/sigir/RenLMR13} focus on the personalized summarization problem
that takes users' interests into account.
Olariu~\cite{DBLP:conf/eacl/Olariu14} devise a graph-based approach to abstractive social summarization.
Bian et al.~\cite{DBLP:journals/tmm/BianYZC15} study the multimedia summarization problem on social streams.
Ren et al.~\cite{DBLP:conf/cikm/RenIAR16} investigate the multi-view opinion summarization of social streams.
Agarwal and Ramamritham~\cite{DBLP:conf/edbt/AgarwalR17} propose a graph-based method for contextual summarization
of social event streams.
Nguyen et al.~\cite{Nguyen:2017:Retaining} consider maintaining a sketch for a social stream
to best preserve the latent topic distribution.

However, the above approaches cannot be applied to ad-hoc query processing
because they (1) do not provide the query interface and (2) are not efficient enough.
For each query, they need to filter out irrelevant elements and
invoke a new instance of the summarization algorithm to acquire the result,
which often takes dozens of seconds or even minutes.
Therefore, it is unrealistic to deploy a summarization method
on a social platform for ad-hoc queries
since thousands of users could submit different queries
at the same time and each query should be processed in real-time.

\textbf{Submodular Maximization.}
Submodular maximization has attracted a lot of research interest recently
for its theoretical significance and wide applications.
The standard approaches to submodular maximization with a cardinality
constraint are the greedy heuristic~\cite{Nemhauser:1978:Analysis}
and its improved version CELF~\cite{Leskovec:2007:Cost},
both of which are $(1-\frac{1}{e})$-approximate.
Badanidiyuru and Vondrak~\cite{Badanidiyuru:2014:FAM}
propose several approximation algorithms for submodular maximization
with general constraints.
Kumar et al.~\cite{Kumar:2015:Fast} and
Badanidiyuru et al.~\cite{Badanidiyuru:2014:Streaming}
study the submodular maximization problem in the distributed and streaming settings.
Epasto et al.~\cite{DBLP:conf/www/EpastoLVZ17} and Wang et al.~\cite{Wang:2018:ERS}
further investigate submodular maximization in the sliding window model.
However, the above algorithms do not utilize any indices for acceleration
and thus they are much less efficient for \query processing than \mtts and \mttd proposed in this paper. 

\section{Problem Formulation}
\label{sec:problem}

\begin{table}
  \centering
  \small
  \caption{Example for social stream and topic model}
  \label{tbl:example:definition}
  \subfloat[Elements extracted from tweets in Figure~\ref{fig:motivation:tweets}]{
    \centering
    \begin{tabular}{|c|c|c|c|c|c|}
    \hline
    \textbf{Elem ID} & \textbf{Time} & \textbf{Words} & $\theta_1$ & $\theta_2$ & \textbf{References} \\
    \hline
    $e_1$ & 1 & $w_1,w_6,w_8,w_{14},w_{16}$ & 0.2  & 0.8  & $\varnothing$ \\ \hline
    $e_2$ & 2 & $w_4,w_9,w_{11}$            & 0.26 & 0.74 & $\varnothing$ \\ \hline
    $e_3$ & 3 & $w_3,w_5,w_{10},w_{13}$     & 0.89 & 0.11 & $\varnothing$ \\ \hline
    $e_4$ & 4 & $w_7,w_{10}$                & 1    & 0    & $e_3$         \\ \hline
    $e_5$ & 5 & $w_6,w_8,w_{16}$            & 0.29 & 0.71 & $e_1$         \\ \hline
    $e_6$ & 6 & $w_2,w_7,w_{10},w_{12}$     & 0.7  & 0.3  & $e_3$         \\ \hline
    $e_7$ & 7 & $w_4,w_{11}$                & 0.33 & 0.67 & $e_2$         \\ \hline
    $e_8$ & 8 & $w_{10},w_{11},w_{15}$      & 0.51 & 0.49 & $e_2,e_3,e_6$ \\
    \hline
    \end{tabular}
  } \\
  \subfloat[Topic-Word distribution -- I]{
    \centering
    \begin{tabular}{|c|c|c|c|}
    \hline
    \textbf{Word ID} & \textbf{Word} & $\theta_1$ & $\theta_2$ \\ \hline
    $w_1$    & asroma   & 0    & 0.03 \\ \hline
    $w_2$    & assist   & 0.06 & 0.04 \\ \hline
    $w_3$    & cavs     & 0.09 & 0    \\ \hline
    $w_4$    & champion & 0.1  & 0.09 \\ \hline
    $w_5$    & defeat   & 0.05 & 0.04 \\ \hline
    $w_6$    & final    & 0.11 & 0.12 \\ \hline
    $w_7$    & lebron   & 0.12 & 0    \\ \hline
    $w_8$    & lfc      & 0    & 0.06 \\ \hline
    \end{tabular}
  }
  \subfloat[Topic-Word distribution -- II]{
    \centering
    \begin{tabular}{|c|c|c|c|}
    \hline
    \textbf{Word ID} & \textbf{Word} & $\theta_1$ & $\theta_2$ \\ \hline
    $w_9$    & manutd      & 0    & 0.07 \\ \hline
    $w_{10}$ & nbaplayoffs & 0.11 & 0    \\ \hline
    $w_{11}$ & pl          & 0    & 0.11 \\ \hline
    $w_{12}$ & point       & 0.15 & 0.14 \\ \hline
    $w_{13}$ & raptors     & 0.08 & 0    \\ \hline
    $w_{14}$ & realmadrid  & 0    & 0.07 \\ \hline
    $w_{15}$ & schedule    & 0.13 & 0.12 \\ \hline
    $w_{16}$ & ucl         & 0    & 0.11 \\ \hline
    \end{tabular}
  }
\end{table}

\subsection{Data Model}
\label{subsec:background}

\textbf{Social Element.}
A social element $e$ is represented as a triple $\langle ts,doc,ref \rangle$,
where $e.ts$ is the timestamp when $e$ is posted,
$e.doc$ is the textual content of $e$ denoted by a bag of words
drawn from a vocabulary $\mathcal{V}$ indexed by $\{1,\ldots,m\}$ ($m=|\mathcal{V}|$),
and $e.ref$ is the set of elements referred to by $e$.
Given two elements $e$ and $e'$ ($e'.ts < e.ts$), if $e$ refers to $e'$,
i.e., $e' \in e.ref$,
we say $e'$ influences $e$, which is denoted as $e' \leadsto e$.
In this way, the attribute $ref$ captures the influence relationships
between social elements~\cite{Subbian:2016:QTI,Wang:2017:RIM}.
If $e$ is totally original, we set $e.ref=\varnothing$.
For example, tweets on Twitter shown in Table~\ref{tbl:example:definition}
are typical social elements and the propagation of hashtags
can be modeled as references~\cite{Subbian:2016:QTI,Li:2017:DYS}.
Note that the \emph{influence relationships} vary for different types of elements,
e.g., ``cite'' between academic papers and ``comment'' on Reddit
can also be modeled as references.

\textbf{Social Stream.}
We consider social elements arrive continuously as a data stream.
A social stream $E$ comprises a sequence
of elements indexed by $\{1,2,3,\ldots\}$.
Elements are ordered by timestamps and multiple elements
with the same timestamp may arrive in an arbitrary manner.
Furthermore, social streams are time-sensitive:
elements posted or referred to recently are more important
and interesting to users than older ones.
To capture the \emph{freshness} of social streams,
we adopt the well-recognized time-based sliding
window~\cite{Datar:2002:Maintaining} model.
Given the window length $T$, a sliding window $W_t$ at time $t$
comprises the elements from time $t-T+1$ to $T$,
i.e., $W_t = \{e \in E | e.ts \in [t-T+1,t]\}$.
The set of active elements $A_t$ at time $t$
includes not only the elements in $W_t$ but also
the elements referred to by any element in $W_t$,
i.e., $A_t = W_t \cup \{e' \in E | e \in W_t \wedge e' \in e.ref\}$.
We use $n_t=|A_t|$ to denote the number of active elements at time $t$.

\textbf{Topic Model.}
We use probabilistic topic models~\cite{Blei:2012:PTM}
such as LDA~\cite{Blei:2003:LDA} and BTM~\cite{DBLP:conf/www/YanGLC13}
to measure the (semantic and influential) representativeness of elements
and the preferences of users.
A topic model $\Theta=\{\theta_{1},\ldots,\theta_{z}\}$ consisting of $z$ topics is
trained from the corpus $\mathcal{E}=\{e.doc | e \in E\}$ and the vocabulary $\mathcal{V}$.
Each topic $\theta_i$ is a multinomial distribution over the words in $\mathcal{V}$,
where $p_{i}(w)$ is the probability of a word $w$ distributed on $\theta_{i}$
and $\sum_{w \in \mathcal{V}} p_{i}(w) = 1$.
The topic distribution of an element $e$ is a multinomial distribution
over the topics in $\Theta$,
where $p_{i}(e)$ is the probability that $e.doc$ is generated from $\theta_i$
and $\sum_{i=1}^{z} p_{i}(e) = 1$.

The selection of appropriate topic models is orthogonal to our problem.
In this work, we consider any probabilistic topic model can be used as a black-box oracle
to provide $p_{i}(w), \forall w \in \mathcal{V}$ and $p_{i}(e), \forall e \in E$.
Note that the evolution of topic distribution is typically much slower than
the speed of social stream~\cite{DBLP:conf/ecir/ZhaoJWHLYL11,DBLP:conf/www/YanGLC13}.
In practice, we assume that the topic distribution remains stable for a period of time.
We need to retrain the topic model from recent elements
when it is outdated due to concept drift.

\subsection{Query Definition}
\label{subsec:define:problem}

\textbf{Query Vector.}
Given a topic model $\Theta$ of $z$ topics,
we use a $z$-dimensional vector $\mathbf{x}=\{x_1,\ldots,x_z\}$ to
denote a user's preference on topics.
Formally, $\mathbf{x} \in [0,1]^{z}$ and,
$x_{i}$ indicates the user's degree of interest on $\theta_{i}$.
W.l.o.g., $\mathbf{x}$ is normalized to $\sum_{i=1}^{z}x_i=1$.
Since it is impractical for users to provide the query vectors directly
for their lack of knowledge about the topic model $\Theta$,
we design a scheme to transform the standard
\emph{query-by-keyword}~\cite{Li:2015:Real} paradigm in our case:
the keywords provided by a user is treated as
a pseudo-document and the query vector is inferred
from its distribution over the topics in $\Theta$.
Note that other query paradigms can also be supported,
e.g., the \emph{query-by-document}~\cite{Zhang:2017:PLQ} paradigm where
a document is provided as a query and the \emph{personalized search}~\cite{Li:2016:CAR}
where the query vector is inferred from a user's recent posts.

\textbf{Definition of Representativeness.}
Given a set of elements $S$ and a query vector $\mathbf{x}$,
the \emph{representativeness} of $S$ w.r.t.~$\mathbf{x}$ at time $t$
is defined by a function $f(\cdot,\cdot) : 2^{|E|} \times [0,1]^z
\rightarrow \mathbb{R}_{\geq 0}$ that maps any subset of $E$
to a nonnegative score w.r.t. a query vector.
Formally, we have
\begin{equation}\label{eq:representative}
f(S,\mathbf{x}) = \sum_{i=1}^{z} x_{i} \cdot f_{i}(S)
\end{equation}
where $f_{i}(S)$ is the score of $S$ on topic $\theta_{i}$.
Intuitively, the overall score of $S$ w.r.t.~$\mathbf{x}$ is the weighted sum
of its scores on each topic.
The score $f_{i}(S)$ on $\theta_i$ is defined
as a linear combination of its semantic and influence
scores. Formally,
\begin{equation}\label{eq:topic-utility}
f_{i}(S) = \lambda \cdot \mathcal{R}_{i}(S) + \frac{1-\lambda}{\eta} \cdot \mathcal{I}_{i,t}(S)
\end{equation}
where $\mathcal{R}_{i}(S)$ is the semantic score of $S$ on $\theta_i$,
$\mathcal{I}_{i,t}(S)$ is the influence score of $S$ on $\theta_i$ at time $t$,
$\lambda \in [0,1]$ specifies the trade-off
between semantic and influence scores,
and $\eta > 0$ adjusts the ranges of
$\mathcal{R}_{i}(\cdot)$ and $\mathcal{I}_{i,t}(\cdot)$ to the same scale.
Next, we will introduce how to compute the semantic and influence
scores based on the topic model $\Theta$ respectively.

\textbf{Topic-specific Semantic Score.}
Given a topic $\theta_i$, we define the semantic score of a set of elements
by the \emph{weighted word coverage} model.
We first define the weight of a word $w$ in $e.doc$ on $\theta_i$.
According to the generative process of topic models~\cite{Blei:2012:PTM},
the probability $p_{i}(w,e)$ that $w \in e.doc$ is generated from $\theta_i$
is denoted as $p_{i}(w,e) = p_{i}(w) \cdot p_{i}(e)$.
Following~\cite{Nguyen:2017:Retaining,Zhuang:2016:DSS},
the weight $\sigma_{i}(w,e)$ of $w$ in $e.doc$ on $\theta_i$ can be defined by
its frequency and information entropy,
i.e., $\sigma_{i}(w,e) = -\gamma(w,e) \cdot p_{i}(w,e) \cdot \log p_{i}(w,e)$,
where $\gamma(w,e)$ is the frequency of $w$ in $e.doc$.
Then, the semantic score of $e$ on $\theta_i$
is the sum of the weights of distinct words in $e.doc$,
i.e., $\mathcal{R}_{i}(e)=\sum_{w \in V_e}\sigma_{i}(w,e)$
where $V_e$ is the set of distinct words in $e.doc$.
We extend the definition of semantic score
to an element set by handling the word overlaps.
Given a set $S$ and a word $w$,
if $w$ appears in more than one element of $S$,
its weight is computed only once for the element $e$
with the maximum $\sigma_{i}(w,e)$.
Formally, the semantic score of $S$ on $\theta_i$ is defined by
\begin{equation}\label{eq:represent-set}
\mathcal{R}_{i}(S) = \sum_{w \in V_S} \max_{e \in S} \sigma_{i}(w,e)
\end{equation}
where $V_S = \cup_{e \in S}V_e$.
Equation~\ref{eq:represent-set} aims to select a set of elements
to maximally cover the important words on $\theta_i$
so as to best preserve the information of $\theta_i$.
Additionally, it implicitly
captures the diversity issue because
adding highly similar elements to $S$ brings
little increase in $\mathcal{R}_{i}(S)$.

\begin{example}\label{exm:semantic:score}
Table~\ref{tbl:example:definition} gives a social stream
extracted from the tweets in Figure~\ref{fig:motivation:tweets}
and a topic model on the vocabulary of elements in the stream.
We demonstrate how to compute the semantic score
$\mathcal{R}_{2}(S)$ where $S=\{e_2,e_7\}$ on $\theta_2$.
The frequency of each word in any element is $1$.
The set of words in $S$ is $V_S=\{w_4,w_9,w_{11}\}$.
The word $w_9$ only appears in $e_2$. Its weight
is $\sigma_{2}(w_9,e_2)=0.15$.
The words $w_4,w_{11}$ appear in both elements.
As $\sigma_{2}(w_4,e_2)=0.18>\sigma_{2}(w_4,e_7)=0.17$
and $\sigma_{2}(w_{11},e_2)=0.20>\sigma_{2}(w_{11},e_7)=0.19$,
$\sigma_{2}(w_4,e_2)$ and $\sigma_{2}(w_{11},e_2)$ are
the weights of $w_4$ and $w_{11}$ for $S$.
Finally, we sum up the weights of each word in $V_S$
and get $\mathcal{R}_{2}(S)=0.53$.
In this example, $e_7$ has no contribution to the semantic score
because all words in $e_7$ are covered by $e_2$.
\end{example}

\textbf{Topic-specific Time-critical Influence Score.}
Given a topic $\theta_i$ and two elements $e',e \in E$ ($e' \in e.ref$),
the probability of influence propagation from $e'$ to $e$ on $\theta_i$
is defined by $p_{i}(e' \leadsto e) = p_{i}(e') \cdot p_{i}(e)$.
Furthermore, the probability of influence propagation from a set of elements $S$
to $e$ on $\theta_i$ is defined by
$p_{i}(S \leadsto e)=1-\prod_{e' \in S \cap e.ref}\big( 1 - p_{i}(e' \leadsto e) \big)$.
We assume the influences from different precedents to $e$ are independent of each other
and adopt the \emph{probabilistic coverage} model to compute the influence probability
from a set of elements to an element.
To select recently trending elements,
we define the influence score in the sliding window model
where only the  references observed within $W_t$ are considered.
Let $I_{t}(e') = \{ e | e' \in e.ref \wedge e \in W_t\}$ be the set of elements
influenced by $e'$ at time $t$
and $I_{t}(S)=\cup_{e' \in S}I_{t}(e')$ be the set of elements
influenced by $S$ at time $t$.
The influence score of $S$ on $\theta_i$ at time $t$ is defined by
\begin{equation}\label{eq:infl-set}
\mathcal{I}_{i,t}(S)=\sum_{e \in I_{t}(S)} p_{i}(S \leadsto e)
\end{equation}
Equation~\ref{eq:infl-set} tends to select a set of influential elements
on $\theta_i$ at time $t$.
The value of $\mathcal{I}_{i,t}(S)$ will increase greatly
only if an element $e$ is added to $S$ such that $e$ is relevant to $\theta_i$ itself
and $e$ is referred to by many elements on $\theta_i$ within $W_t$.

\begin{example}\label{exm:influence:score}
We compute the influence score $\mathcal{I}_{2,8}(S)$
of $S=\{e_2,e_3\}$ in Table~\ref{tbl:example:definition} on $\theta_2$ at time $t=8$.
We consider the window length $T=4$ and $W_{t}=\{e_5,e_6,e_7,e_8\}$.
$I_{8}(S)$ at time $8$ is $\{e_6,e_7,e_8\}$ and $e_4$ expires at time $8$.
First, $p_{2}(S \leadsto e_6) = p_{2}(e_3 \leadsto e_6) = 0.03$.
Similarly, $p_{2}(S \leadsto e_7) = p_{2}(e_2 \leadsto e_7) = 0.50$.
For $e_8$, we have $p_{2}(S \leadsto e_8) = 1-
\big( 1-p_{2}(e_2 \leadsto e_8) \big) \cdot \big( 1-p_{2}(e_3 \leadsto e_8) \big) = 0.40$.
Finally, we acquire $\mathcal{I}_{2,8}(S) = 0.03 + 0.5 + 0.4 =0.93$.
We can see, although $e_3$ is referred to by several elements,
its influence score on $\theta_2$ is low
because $e_3$ and the elements referring to it
are mostly on $\theta_1$.
\end{example}

\textbf{Query Definition.}
We formally define the \emph{Semantic and Influence aware $k$-Representative}
(\query) query to select a set of elements $S$ with the maximum representativeness
score w.r.t.~a query vector from a social stream.
We have two constraints on the result of \query query $S$:
(1) its size is restricted to $k \in \mathbb{Z}^+$,
i.e., $S$ contains at most $k$ elements,
to avoid overwhelming users with too much information;
(2) the elements in $S$ must be active at time $t$, i.e., $S \subseteq A_t$,
to satisfy the freshness requirement.
Finally, we define a \query query $q_t(k,\mathbf{x})$ as follows.

\begin{definition}[\query]\label{def:query}
Given the set of active elements $A_t$ and a vector $\mathbf{x}$,
a \query query $q_t(k,\mathbf{x})$ returns a set of elements $S^{*} \subseteq A_t$
with a bounded size $k$ such that the scoring function $f(\cdot,\mathbf{x})$
is maximized, i.e., $S^{*} = \argmax_{S \subseteq A_t : |S| \leq k} f(S,\mathbf{x})$,
where $S^{*}$ is the optimal result for $q_t(k,\mathbf{x})$
and $\mathtt{OPT} = f(S^{*},\mathbf{x})$ is the optimal representativeness score.
\end{definition}

\begin{example}\label{exm:query}
We consider two \query queries on the social stream in Table~\ref{tbl:example:definition}.
We set $\lambda=0.5$, $\eta=2$ in Equation~\ref{eq:topic-utility} and the window length $T=4$.
At time $8$, the set of active elements $A_t$ contains all except $e_4$.
Given a \query query $q_8(2,\mathbf{x}_1)$ where $\mathbf{x}_1=(0.5,0.5)$
(a user has the same interest on two topics),
$S^{*}=\{e_1,e_3\}$ is the query result
and $\mathtt{OPT}=f(S^{*},\mathbf{x}_1)=0.65$.
We can see $e_3,e_1$ obtain the highest scores
on $\theta_1,\theta_2$ respectively
and they collectively achieve the maximum score w.r.t.~$\mathbf{x}_1$.
Given an \query query $q_8(2,\mathbf{x}_2)$ where $\mathbf{x}_2=(0.1,0.9)$
(the user prefers $\theta_2$ to $\theta_1$),
the query result is $S^{*}=\{e_1,e_2\}$ and $\mathtt{OPT}=0.94$.
$e_3$ is excluded because it is mostly distributed on $\theta_1$.
\end{example}

\subsection{Properties and Challenges}

\textbf{Properties of \query Queries.}
We first show the \emph{monotonicity} and \emph{submodularity} of the scoring
function $f(\cdot,\cdot)$ for \query query by proving that both
the semantic function $\mathcal{R}_{i}(\cdot)$
and the influence function $\mathcal{I}_{i,t}(\cdot)$
are monotone and submodular.

\begin{definition}[Monotonicity \& Submodularity]
A function $g(\cdot): 2^{|E|} \rightarrow \mathbb{R}_{\geq 0}$ on the power set of $E$
is monotone iff $g(S\cup\{e\}) \geq g(S)$ for any $e \in E \setminus S$ and $S \subseteq E$.
The function $g(\cdot)$ is submodular iff $g(S\cup\{e\})-g(S) \geq g(T\cup\{e\})-g(T)$
for any $S \subseteq T \subseteq E$ and $e \in E \setminus T$.
\end{definition}

\begin{lemma}\label{prop:semantic:submodular}
$\mathcal{R}_{i}(\cdot)$ is monotone and submodular for $i \in [1,z]$.
\end{lemma}
\begin{lemma}\label{prop:influence:submodular}
$\mathcal{I}_{i,t}(\cdot)$ is monotone and submodular for $i \in [1,z]$ at any time $t$.
\end{lemma}
The proofs are given in Appendices~\ref{subsec:proof:lemma:1} and~\ref{subsec:proof:lemma:2}.

Given a query vector $\mathbf{x}$, the scoring function
$f(\cdot,\mathbf{x})$ is a nonnegative linear combination
of $\mathcal{R}_{i}(\cdot)$ and $\mathcal{I}_{i,t}(\cdot)$.
Therefore, $f(\cdot,\mathbf{x})$ is \emph{monotone} and \emph{submodular}.

\textbf{Challenges of \query Queries.}
In this paper, we consider that the elements arrive continuously over time.
We always maintain the set of active elements $A_t$ at any time $t$.
It is required to provide the result for any ad-hoc \query query $q_t(k,\mathbf{x})$ in real-time.

The challenges of processing \query queries in such a scenario are two-fold:
(1) \emph{NP-hardness} and (2) \emph{dynamism}.
First, the following theorem shows the \query query is \emph{NP-hard}.
\begin{theorem}\label{thm:nphard}
It is NP-hard to obtain the optimal result $S^{*}$ for any \query query $q_t(k,\mathbf{x})$.
\end{theorem}
The \emph{weighted maximum coverage} problem can be reduced to \query query
when $\lambda=1$ in Equation~\ref{eq:topic-utility}.
Meanwhile, the \emph{probabilistic coverage} problem
is a special case of \query query when $\lambda=0$ in Equation~\ref{eq:topic-utility}.
Because both problems are NP-hard~\cite{Feige:1998:TLN},
the \query query is NP-hard as well.

In spite of this, existing algorithms for submodular
maximization~\cite{Nemhauser:1978:Analysis} can provide
results with constant approximations to the optimal ones for \query queries
due to the monotonicity and submodularity of the scoring function.
For example, CELF~\cite{Leskovec:2007:Cost} is $(1-\frac{1}{e})$-approximate
for \query queries while SieveStreaming~\cite{Badanidiyuru:2014:Streaming}
is $(\frac{1}{2}-\varepsilon)$-approximate (for any $\varepsilon > 0$).
However, both algorithms cannot fulfill the requirements for
real-time \query processing owing to the dynamism of \query queries.
The results of \query queries not only vary with query vectors but also
evolve over time for the same query vector
due to the changes in active elements
and the fluctuations in influence scores over the sliding window.
To process one \query query $q_t(k,\mathbf{x})$, CELF
and SieveStreaming should evaluate
$f(\cdot,\mathbf{x})$ for $\mathcal{O}(k \cdot n_t)$
and $\mathcal{O}(\frac{\log k}{\varepsilon} \cdot n_t)$ times respectively.
Empirically, they often take several seconds for one \query query when the window length is 24 hours.
To the best of our knowledge, none of the existing algorithms can efficiently process \query queries.
Thus, we are motivated to devise novel real-time solutions for \query processing over social streams.

\begin{table}
  \centering
  \caption{Frequently Used Notations}
  \label{tbl:notations}
  \begin{tabular}{l|m{5.35in}}
    \hline
    \textbf{Notation} & \textbf{Description} \\
    \hline
    $E,e,e_{i}$ & $E=\{e_1,\ldots,e_n\}$ is a social stream;
    $e$ is an arbitrary element in $E$; $e_{i}$ is the $i$-th element in $E$.\\ \hline
    $T,W_t,A_t$ & $T$ is the window length; $W_t$ is the sliding window at time $t$;
    $A_t$ is the set of active elements at time $t$.\\ \hline
    $\Theta,\theta_{i}$ & $\Theta$ is a topic model; $\theta_{i}$ is the $i$-th topic in $\Theta$.\\ \hline
    $\mathbf{x}, x_i$ & $\mathbf{x}$ is a $z$-dimensional vector; $x_i$ is the $i$-th entry of $\mathbf{x}$.\\ \hline
    $\mathcal{R}_{i}(\cdot), \mathcal{I}_{i,t}(\cdot)$ & $\mathcal{R}_{i}(\cdot)$ is the semantic function on $\theta_{i}$; $\mathcal{I}_{i,t}(\cdot)$ is the influence function on $\theta_{i}$ at time $t$.\\ \hline
    $f_i(\cdot), f(\cdot,\cdot)$ & $f_i(\cdot)$ is the representativeness scoring function on $\theta_{i}$;
    $f(\cdot,\cdot)$ is the scoring function w.r.t. a query vector.\\ \hline
    $q_t(k,\mathbf{x})$ & $q_t(k,\mathbf{x})$ is a \query query at time $t$ with a bounded result size $k$ and a query vector $\mathbf{x}$.\\ \hline
    $S^{*},\mathtt{OPT}$ & $S^{*}$ is the optimal result for $q_t(k,\mathbf{x})$;
    $\mathtt{OPT}=f(S^{*},\mathbf{x})$ is the optimal representativeness score.\\ \hline
    $\delta_{i}(e),\delta(e,\mathbf{x})$ & $\delta_{i}(e)=f_i(\{e\})$ is the score of $e$ on $\theta_{i}$;
    $\delta(e,\mathbf{x})=f(\{e\},\mathbf{x})$ is the score of $e$ w.r.t.~$\mathbf{x}$.\\ \hline
    $\Delta(e|S)$ & $\Delta(e|S)=f(S\cup\{e\},\mathbf{x})-f(S,\mathbf{x})$ is the marginal score gain of adding $e$ to $S$.\\ \hline
    $\rl_{i}$ & $\rl_{i}$ is the ranked list maintained for the elements on topic $\theta_{i}$.\\
    \hline
  \end{tabular}
\end{table}

Before moving on to the section for \query processing,
we summarize the frequently used notations in Table~\ref{tbl:notations}.

\section{Query Processing}
\label{sec:framework}

\begin{figure}
  \centering
  \includegraphics[width=0.6\textwidth]{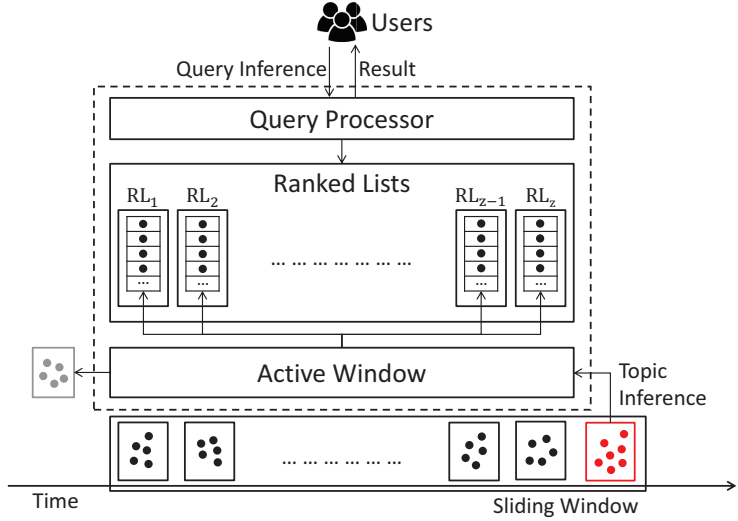}
  \caption{The architecture for \query query processing}
  \label{fig:framework}
\end{figure}

In this section, we introduce the methods to process \query queries over social streams.
The architecture is illustrated in Figure~\ref{fig:framework}.
At any time $t$, we maintain (1) \textbf{Active Window} to
buffer the set of active elements $A_t$,
(2) \textbf{Ranked Lists} $\rl_{1},\ldots,\rl_{z}$ to sort the lists of elements
on each topic of $\Theta$ in descending order of topic-wise representativeness score,
and (3) \textbf{Query Processor} to
leverage the \textit{ranked lists} to process \query queries.
In addition, when the topic model is given,
the query and topic inferences become rather standard
(e.g., Gibbs sampling~\cite{Liu:2011:PPL}),
and thus we do not discuss these procedures here for space limitations.
We consider the query vectors and the topic vectors of elements have been given in advance.

As shown in Figure~\ref{fig:framework}, we process a social stream $E$ in a batch manner.
$E$ is partitioned into buckets with equal time length $L\in\mathbb{Z}^{+}$
and updated at discrete time $L,2L,\ldots$ until the end time of the stream $t_n$.
When the window slides at time $t$, a bucket $B_t$
containing the elements between time $t-L+1$ to $t$ is received.
After inferring the topic vector of each $e \in B_t$ with the topic model,
we first update the \textit{active window}.
The elements in $B_t$ are inserted into the \textit{active window} and
the elements referred to by them are updated.
Then, the elements that are never referred to by any element
after time $t-T+1$ are discarded from the \textit{active window}.
Subsequently, the \textit{ranked list} $\rl_{i}$
on each topic $\theta_{i}$ is maintained for $B_t$.
The detailed procedure for ranked lists maintenance
will be presented in Section~\ref{subsec:rankedlist}.

Next, let us discuss the mechanism of \query processing.
One major drawback of existing submodular maximization methods,
e.g., CELF~\cite{Leskovec:2007:Cost} and SieveStreaming~\cite{Badanidiyuru:2014:Streaming},
on processing \query queries is that they
need to evaluate every active element at least once.
However, real-world datasets often have two characteristics:
(1) The scores of elements are skewed,
i.e., only a few elements have high scores.
For example, we compute the scores of a sample of tweets w.r.t. a \query query
and scale the scores linearly to the range of 0 to 1.
The statistics demonstrate that
only 0.4\% elements have scores of greater than 0.9
while 91\% elements have scores of less than 0.1.
(2) One element can only be high-ranked in very few topics,
i.e., one element is about only one or two topics.
In practice, we observe that the average number of topics per element
is less than 2.
Therefore, most of the elements are not relevant to a specific \query query.
We can greatly improve the efficiency by avoiding the evaluations for the elements
with very low chances to be included into the query result.
To prune these unnecessary evaluations, we leverage the ranked lists
to sequentially evaluate the active elements in decreasing order
of their scores w.r.t. the query vector.
In this way, we can track whether unevaluated elements
can still be added to the query result
and terminate the evaluations as soon as possible.

Although such a method to traverse the \textit{ranked lists}
is similar to the one for top-$k$ query~\cite{Zhang:2017:PLQ},
the procedures for maintaining the query results are totally different.
A top-$k$ query simply returns $k$ elements with the maximum
scores as the result for a \query query.
Although the top-$k$ result can be retrieved
efficiently from the \textit{ranked lists}
using existing methods~\cite{Zhang:2017:PLQ},
its quality for \query queries is suboptimal
because the word and influence overlaps are ignored.
Thus, we will propose the \textsc{Multi-Topic ThresholdStream} (\mtts)
and \textsc{Multi-Topic ThresholdDescend} (\mttd) algorithms for \query processing
in Sections~\ref{subsec:mtts} and~\ref{subsec:mttd}.
They can return high-quality results with constant approximation guarantees
for \query queries while meeting the real-time requirements.

\subsection{Ranked List Maintenance}
\label{subsec:rankedlist}

In this subsection, we introduce the procedure for ranked list maintenance.
Generally, a ranked list $\rl_{i}$ keeps a tuple for each active element on topic $\theta_i$.
A tuple for element $e$ is denoted as $\langle \delta_{i}(e),t_e \rangle$
where $\delta_{i}(e)=f_i(\{e\})$ is the topic-wise representativeness score of $e$ on $\theta_i$
and $t_e$ is the timestamp when $e$ is last referred to.
All tuples in $\rl_{i}$ are sorted in descending order of topic-wise score.

\begin{algorithm}[t]
  \KwIn{A social stream $E$, the window length $T$, and the bucket length $L$}
  $t \gets 0$, initialize an empty ranked list $\rl_{i}$ for $i\in[1,z]$\;\label{line:rl:initial}
  \While{$t \leq t_n$}{
    $t \gets t+L, B_{t} \gets \{e \in E | e.ts\in[t-L+1,t]\}$\;\label{line:ranked:list:prepare}
    \ForEach{$e \in B_{t}$\label{line:ranked:list:doc:start}}{
      \ForEach{$i : p_{i}(e)>0$}{
        $\delta_{i}(e) \gets \mathcal{R}_{i}(e), t_e \gets e.ts$\;
        create a tuple $\langle \delta_{i}(e),t_e \rangle$ and insert it into $\rl_{i}$\;\label{line:ranked:list:doc:end}
      }
      \ForEach{$e' \in e.ref$\label{line:ranked:list:ref:start}}{
        \ForEach{$i : p_{i}(e')>0 \wedge p_{i}(e)>0$}{
          $\delta_{i}(e') \gets f_i(\{e'\}), t_{e'} \gets e.ts$\;
          adjust the position of $\langle \delta_{i}(e'),t_{e'} \rangle$ in $\rl_{i}$\;\label{line:ranked:list:ref:end}
        }
      }
    }
    \ForEach{$e$ : $e$ is never referred to after $t-T+1$\label{line:ranked:list:delete:start}}{
      delete the tuples of $e$ from $\rl_{i}$ with $p_{i}(e)>0$\;\label{line:ranked:list:delete:end}
    }
  }
  \caption{\textsc{Ranked List Maintenance}}
  \label{alg:ranked:list}
\end{algorithm}

The algorithmic description of ranked list maintenance over a social stream
is presented in Algorithm~\ref{alg:ranked:list}.
Initially, an empty ranked list is initialized for each topic $\theta_i$
in the topic model $\Theta$ (Line~\ref{line:rl:initial}).
At discrete timestamps $t=L,2L,\ldots$ until $t_n$,
the \textit{ranked lists} are updated according to a bucket of elements $B_{t}$.
For each element $e$ in $B_{t}$, a tuple $\langle \delta_{i}(e),t_{e} \rangle$
is created and inserted into $\rl_{i}$ for every topic $\theta_{i}$ with $p_{i}(e)>0$
(Lines~\ref{line:ranked:list:doc:start}--\ref{line:ranked:list:doc:end}).
The score $\delta_{i}(e)$ is $\mathcal{R}_{i}(e)$
because the elements influenced by $e$ have not been observed yet.
The time $t_e$ when $e$ is last referred to is obviously $e.ts$.
Subsequently, it recomputes the influence score $\mathcal{I}_{i,t}(e')$
for each parent $e'$ of $e$. After that, it updates the tuple
$\langle \delta_{i}(e'),t_{e'} \rangle$ by setting $\delta_{i}(e')$ to $f_i(\{e'\})$
and $t_{e'}$ to $e.ts$.
The position of $\langle \delta_{i}(e'),t_{e'} \rangle$ in $\rl_{i}$
is adjusted according to the updated $\delta_{i}(e')$
(Lines~\ref{line:ranked:list:ref:start}--\ref{line:ranked:list:ref:end}).
Finally, we delete the tuples for expired elements from $\rl_{i}$
(Lines~\ref{line:ranked:list:delete:start}--\ref{line:ranked:list:delete:end}).

\textbf{Complexity Analysis.}
The cost of evaluating $\delta_{i}(e)$ for any element $e$
is $\mathcal{O}(l)$ where $l=\max_{e \in A_t}$ $(|V_e|+|I_t(e)|)$.
Then, the complexity of inserting a tuple into $\rl_{i}$ is $\mathcal{O}(\log n_{t})$.
For each $e' \in e.ref$, the complexity of re-evaluating $\mathcal{I}_{i,t}(e')$
is also $\mathcal{O}(l)$.
Overall, the complexity of maintaining $\rl_{i}$ for element $e$
is $\mathcal{O}\big(\mathcal{P} (l+\log n_{t})\big)$
where $\mathcal{P}=\max_{e \in A_t}|e.ref|$.
As the tuples for $e$ may appear in $\mathcal{O}(z)$ ranked lists,
the time complexity of ranked list maintenance for element $e$
is $\mathcal{O}\big(z \mathcal{P}(l+\log n_{t})\big)$.

\textbf{Operations for Ranked List Traversal.}
We need to access the tuples in each \textit{ranked list} $\rl_{i}$
in decreasing order of topic-wise score for \query processing.
Two basic operations are defined to traverse the ranked
list $\rl_{i}$: (1) $\rl_{i}.\mathsf{first}$ to retrieve the element
w.r.t.~the first tuple with the maximum topic-wise score from $\rl_{i}$;
(2) $\rl_{i}.\mathsf{next}$ to acquire the element w.r.t.~the next unvisited
tuple in $\rl_{i}$ from the current one. Note that once a tuple for element $e$
has been accessed in one ranked list, the remaining tuples for $e$ in the other lists
will be marked as ``visited'' so as to eliminate duplicate evaluations for $e$.

\subsection{Multi-Topic ThresholdStream Algorithm}
\label{subsec:mtts}

In this subsection, we present the \mtts algorithm for \query processing.
\mtts is built on two key ideas:
(1) a thresholding approach~\cite{Kumar:2015:Fast} to submodular maximization
and (2) a ranked list based mechanism for early termination.
First, given a \query query, the thresholding approach always tracks its optimal
representativeness score $\mathtt{OPT}$. It establishes
a sequence of candidates with different \textit{thresholds}
within the range of $\mathtt{OPT}$. For any element $e$,
each candidate determines whether to include $e$ independently based on $e$'s
marginal gain and its threshold.
Second, to prune unnecessary evaluations,
\mtts utilizes \emph{ranked lists} to sequentially feed elements
to the candidates in decreasing order of score. It continuously
checks the minimum threshold for an element to be added to any candidate
and the upper-bound score of unevaluated elements.
\mtts is terminated when the upper-bound score is lower than the minimum threshold.
After termination, the candidate with the maximum score is returned
as the result for the \query query.

\begin{algorithm}[t]
  \KwIn{The ranked list $\rl_{i}$ for each $i \in [1,z]$ and a \query query $q_t(k,\mathbf{x})$}
  \Parameter{$\varepsilon \in (0,1)$}
  \KwResult{$S_{ts}$ for $q_t(k,\mathbf{x})$}
  $\Phi=\{(1+\varepsilon)^{j}|j \in \mathbb{Z}\}$,
  \textbf{foreach} $\phi \in \Phi$ \textbf{do} $S_{\phi} \gets \varnothing$\;\label{line:mtts:init:start}
  \textbf{foreach} $i \in [1,z] : x_i>0$ \textbf{do} $e^{(i)} \gets \rl_{i}.\mathsf{first}$\;
  $\delta_{max},\mathtt{TH} \gets 0$ and $\mathtt{UB}(\mathbf{x}) \gets \sum_{i=1}^{z} x_i \cdot \delta_{i}(e^{(i)})$\;\label{line:mtts:init:end}
  \While{$\mathtt{UB}(\mathbf{x}) \geq \mathtt{TH}$\label{line:mtts:rankedlist:start}}{
    $i^{*} \gets \argmax_{i \in [1,z]} x_i \cdot \delta_{i}(e^{(i)}), e \gets e^{(i^{*})}$\;\label{line:mtts:next:doc}
    $\delta(e,\mathbf{x}) \gets \sum_{i=1}^{z} x_i \cdot \delta_i(e)$\;\label{line:mtts:cand:start}
    \textbf{if} $\delta(e,\mathbf{x})>\delta_{max}$ \textbf{then} $\delta_{max} \gets \delta(e,\mathbf{x})$\;
    $\Phi=\{(1+\varepsilon)^{j}|j \in \mathbb{Z} \wedge \delta_{max} \leq (1+\varepsilon)^{j} \leq 2 \cdot k \cdot \delta_{max}\}$\;
    delete $S_{\phi}$ if $\phi \notin \Phi$\;\label{line:mtts:cand:end}
    \ForEach{$\phi\in\Phi$\label{line:mtts:eval:start}}{
      \If{$\delta(e,\mathbf{x}) \geq \frac{\phi}{2k} \wedge |S_{\phi}|<k$}{
        \textbf{if} $\Delta(e|S_{\phi}) \geq \frac{\phi}{2k}$
        \textbf{then} $S_{\phi} \gets S_{\phi} \cup \{e\}$\;
      }\label{line:mtts:eval:end}
    }
    $e^{(i^{*})} \gets \rl_{i^{*}}.\mathsf{next}$\;\label{line:mtts:var:next}
    $\mathtt{TH} \gets \min_{\phi\in\Phi : |S_{\phi}| < k} \frac{\phi}{2k}$,
    $\mathtt{UB}(\mathbf{x}) \gets \sum_{i=1}^{z} x_i \cdot \delta_{i}(e^{(i)})$\;\label{line:mtts:var:update}\label{line:mtts:rankedlist:end}
  }
  \Return{$S_{ts} \gets \argmax_{\phi \in \Phi} f(S_{\phi},\mathbf{x})$}\;\label{line:mtts:result}
  \caption{\textsc{Multi-Topic ThresholdStream}}
  \label{alg:mtts}
\end{algorithm}

The algorithmic description of \mtts is presented in Algorithm~\ref{alg:mtts}.
The initialization phase is shown in Lines~\ref{line:mtts:init:start}--\ref{line:mtts:init:end}.
Given a parameter $\varepsilon \in (0,1)$, \mtts establishes a geometric progression
$\Phi$ with common ratio $(1+\varepsilon)$ to estimate the optimal
score $\mathtt{OPT}$ for $q_t(k,\mathbf{x})$. Then, it maintains a candidate $S_{\phi}$
initializing to $\varnothing$ for each $\phi\in\Phi$.
The threshold for $S_{\phi}$ is $\frac{\phi}{2k}$.
The traversal of ranked lists starts from the first tuple of each list.
We use $e^{(i)}$ to denote the element corresponding to the current tuple from $\rl_{i}$.
\mtts keeps $3$ variables: (1) $\delta_{max}$ to store the maximum score w.r.t.~$\mathbf{x}$
among the evaluated elements, (2) $\mathtt{TH}$ to maintain the minimum threshold
for an element to be added to any candidate, and 3) $\mathtt{UB}(\mathbf{x})$
to track the upper-bound score for any unevaluated element w.r.t.~$\mathbf{x}$.
Specifically, $\mathtt{TH}$ is the threshold $\frac{\phi}{2k}$
of the unfilled candidate $S_{\phi}$ (i.e., $|S_{\phi}|<k$) with the minimum $\phi$.
We set $\mathtt{TH}=0$ before the evaluation.
If $\delta(e,\mathbf{x})<\mathtt{TH}$, $e$ can be safely excluded from evaluation.
In addition, for any unevaluated element $e$, it holds that
$\delta_{i}(e) \leq \delta_{i}(e^{(i)})$ because the tuples in $\rl_{i}$
are sorted by topic-wise score.
Thus, $\mathtt{UB}(\mathbf{x}) = \sum_{i=1}^{z} x_i \cdot \delta_{i}(e^{(i)})$ can be used
as the upper-bound score of unevaluated elements w.r.t.~$\mathbf{x}$.

After the initialization phase, the elements are sequentially retrieved
from the ranked lists and evaluated by the candidates according to
Lines~\ref{line:mtts:rankedlist:start}--\ref{line:mtts:rankedlist:end}.
At each iteration, \mtts selects an element $e^{(i^{*})}$
with the maximum $x_i \cdot \delta_{i}(e^{(i)})$
as the next element $e$ for evaluation (Line~\ref{line:mtts:next:doc}).
Subsequently, the candidate maintenance procedure is performed following
Lines~\ref{line:mtts:cand:start}--\ref{line:mtts:cand:end}.
It first computes the score $\delta(e,\mathbf{x})$ of $e$ w.r.t.~$\mathbf{x}$.
Second, it updates the maximum score $\delta_{max}$.
Third, the range of $\mathtt{OPT}$ is
adjusted to $[\delta_{max},2 \cdot k \cdot \delta_{max}]$.
Fourth, it deletes the candidates out of the range for $\mathtt{OPT}$.
Next, each candidate $S_{\phi}$ determines whether to add $e$
independently according to Lines~\ref{line:mtts:eval:start}--\ref{line:mtts:eval:end}.
If $\delta(e,\mathbf{x})<\frac{\phi}{2k}$ or $S_{\phi}$ has contained $k$ elements,
$e$ will be ignored by $S_{\phi}$.
Otherwise, the marginal gain $\Delta(e|S_{\phi})=f(S_{\phi}\cup\{e\},\mathbf{x})-f(S_{\phi},\mathbf{x})$
of adding $e$ to $S_{\phi}$ is evaluated.
If $\Delta(e|S_{\phi})$ reaches $\frac{\phi}{2k}$, $e$ will be added to $S_{\phi}$.
Finally, it obtains the next element in $\rl_{i^{*}}$ as $e^{(i^{*})}$
and updates $\mathtt{TH},\mathtt{UB}(\mathbf{x})$ accordingly
(Lines \ref{line:mtts:var:next} and~\ref{line:mtts:var:update}).
The evaluation procedure will be terminated when $\mathtt{UB}(\mathbf{x})<\mathtt{TH}$
because $\delta(e',\mathbf{x}) \leq \mathtt{UB}(\mathbf{x}) < \mathtt{TH}$
is satisfied for any unevaluated element $e'$,
which can be safely pruned.
Finally, \mtts returns the candidate with the maximum score
as the result for $q_t(k,\mathbf{x})$ (Line~\ref{line:mtts:result}).

\begin{figure}
  \centering
  \includegraphics[width=0.6\textwidth]{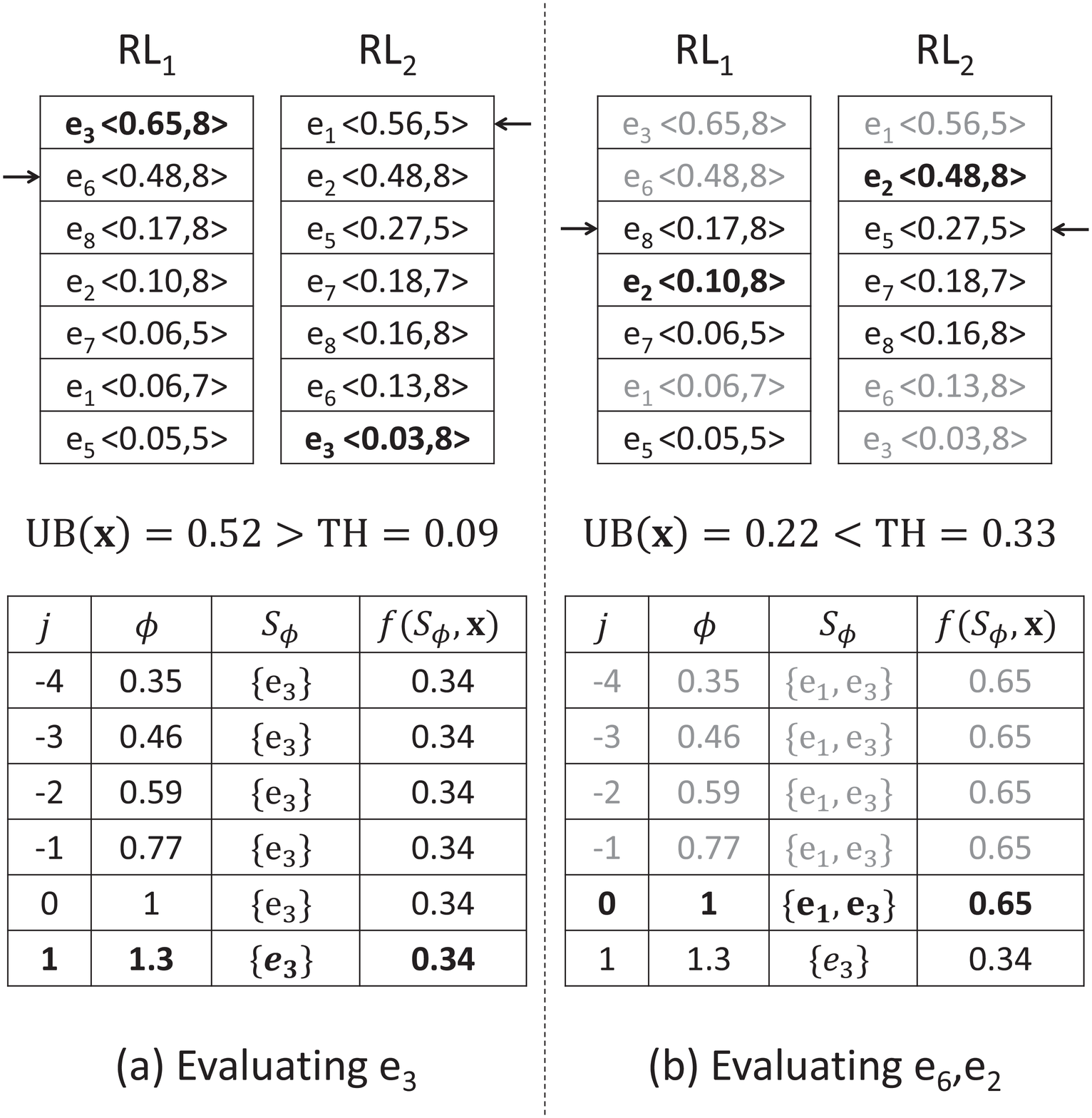}
  \caption{Example for \query processing using \mtts.}
  \label{fig:mtts}
\end{figure}

\begin{example}
Following the example in Table~\ref{tbl:example:definition},
we show how \mtts processes a \query query $q_8(2,\mathbf{x})$
where $\mathbf{x}=(0.5,0.5)$ in Figure~\ref{fig:mtts}.
We set $\varepsilon=0.3$ in this example.

First of all, the traversals of $\rl_{1}$ and $\rl_{2}$ start from $e_3$ and $e_1$ respectively.
Initially, we have $\mathtt{UB}(\mathbf{x})=0.61$ and $\mathtt{TH}=0$.
Then, the first element to evaluate is $e_3$ because
$x_1 \cdot \delta_1(e_3)=0.33 > x_2 \cdot \delta_2(e_1)=0.28$.
As $\delta(e_3,\mathbf{x})=0.34$, the range of $\mathtt{OPT}$
is $[0.34,1.36]$. We have $0.34 \leq {1.3}^{-4} \leq \ldots \leq {1.3}^{1} \leq 1.36$
and $6$ candidates with $j\in[-4,1]$ are maintained.
$e_3$ can be added to each of the candidates.
After that, $e_6$ is the next element from $\rl_{1}$.
$\mathtt{UB}(\mathbf{x})$ and $\mathtt{TH}$ are updated to $0.52$ and $0.09$ respectively.
The second element to evaluate is $e_1$ from $\rl_{2}$.
As $\delta(e_1,\mathbf{x})=0.31$, the candidate with $j=1$ directly skips $e_1$
for $\delta(e_1,\mathbf{x})<\frac{\phi}{2k}=0.33$. Other candidates
include $e_1$ as $\Delta(e_1|S_{\phi})\geq\frac{\phi}{2k}$.
Then, $e_2$ is the next element from $\rl_{2}$.
$\mathtt{UB}(\mathbf{x})$ decreases to $0.48$ while $\mathtt{TH}$ increases to $0.33$.
Subsequently, $e_6,e_2$ are retrieved but skipped by all candidates.
After evaluating $e_2$, $\mathtt{UB}(\mathbf{x})$ decreases to $0.22$ and is lower than $\mathtt{TH}$.
Thus, no more evaluation is needed and $S_{ts}=\{e_1,e_3\}$
is returned as the result for $q_8(2,\mathbf{x})$.
\end{example}

The approximation ratio of \mtts is given in
Theorem~\ref{thm:mtts:ratio}.

\begin{theorem}\label{thm:mtts:ratio}
$S_{ts}$ returned by \mtts is a $(\frac{1}{2}-\varepsilon)$-approximation result for any \query query.
\end{theorem}
The proof is given in Appendix~\ref{subsec:proof:theorem:1}.

\textbf{Complexity Analysis.}
The number of candidates in \mtts is $\mathcal{O}(\frac{\log k}{\varepsilon})$
as the ratio between the lower and upper bounds for $\mathtt{OPT}$ is $\mathcal{O}(k)$.
The complexity of retrieving an element from ranked lists is $\mathcal{O}(\log n_t)$.
The complexity of evaluating one element for a candidate is $\mathcal{O}(ld)$
where $l=\max_{e \in A_t}(|V_e|+|I_t(e)|)$ and $d$ is
the number of non-zero entries in the query vector $\mathbf{x}$.
Thus, the complexity of \mtts to evaluate one element is
$\mathcal{O}(\log n_t + \frac{l d \log k}{\varepsilon})$.
Overall, the time complexity of \mtts is
$\mathcal{O}\big(n_t'(\log n_t + \frac{l d \log k}{\varepsilon})\big)$
where $n_t'$ is the number of elements evaluated by \mtts.

\subsection{Multi-Topic ThresholdDescend Algorithm}
\label{subsec:mttd}

Although \mtts is efficient for \query processing, its approximation ratio is lower than
the the best achievable approximation guarantees, i.e., $(1-\frac{1}{e})$~\cite{Feige:1998:TLN}
for submodular maximization with cardinality constraints.
In addition, its result quality is also slightly inferior to that of CELF.
In this subsection, we propose the \textsc{Multi-Topic ThresholdDescend} (\mttd)
algorithm to improve upon \mtts.
Different from \mtts, \mttd maintains only one candidate
$S$ from $\varnothing$ to reduce the cost for evaluation.
In addition, it buffers the elements that are retrieved from ranked lists
but not included into $S$ so that these elements can be evaluated more than once.
This can lead to better quality as the chances of missing significant
elements are smaller.
Specifically, \mttd has multiple rounds of evaluation with decreasing \emph{thresholds}.
In the round with threshold $\tau$, each element $e$
with $\delta(e,\mathbf{x}) \geq \tau$ is considered
and will be included to $S$ once the marginal gain $\Delta(e|S)$ reaches $\tau$.
When $S$ contains $k$ elements or $\tau$ is descended to the lower bound,
\mttd is terminated and $S$ is returned as the result.
Theoretically, the approximation ratio of \mttd
is improved to $(1-\frac{1}{e}-\varepsilon)$
but its worst-case complexity is higher than \mtts.
Despite this, the efficiency and result quality of \mttd
are both better than \mtts empirically.

\begin{algorithm}[t]
  \SetKwFunction{retrieve}{retrieve}
  \KwIn{The ranked list $\rl_{i}$ for each $i \in [1,z]$ and a \query query $q_t(k,\mathbf{x})$}
  \Parameter{$\varepsilon \in (0,1)$}
  \KwResult{$S_{td}$ for $q_t(k,\mathbf{x})$}
  $S,E' \gets \varnothing$\;\label{line:mttd:init:start}
  \textbf{foreach} $i \in [1,z] : x_i>0$ \textbf{do} $e^{(i)} \gets \rl_{i}.\mathsf{first}$\;
  $\tau \gets \sum_{i=1}^{z} x_i \cdot \delta_{i}(e^{(i)}),\tau' \gets 0$\;\label{line:mttd:init:end}
  \While{$\tau \geq \tau'$\label{line:mttd:round:start}}{
    $E_{\tau} \gets$ \retrieve{$\tau$}, $E' \gets E' \cup E_{\tau}$\;
    \While{$\exists e \in E' \setminus S : \Delta_{e} \geq \tau$\label{line:mttd:eval:start}}{
      $e' \gets \argmax_{e \in E' \setminus S} \Delta_{e}, \Delta_{e'} \gets \Delta(e'|S)$\;
      \If{$\Delta_{e'} \geq \tau$}{
        $S \gets S \cup \{e'\}, E' \gets E' \setminus \{e'\}$\;
        \textbf{if} $|S|=k$ \textbf{then} \Return{$S_{td} \gets S$}\;
      }\label{line:mttd:eval:end}
    }
    $\tau' \gets f(S,\mathbf{x})\cdot\frac{\varepsilon}{k}, \tau \gets (1-\varepsilon)\tau$\;\label{line:mttd:round:end}
  }
  \Return{$S_{td} \gets S$}\;\label{line:mttd:result}
  \BlankLine
  \SetKwProg{proc}{Procedure}{}{}
  \proc{\retrieve{$\tau$}\label{line:mttd:retrieve:start}}{
    $E_{\tau} \gets \varnothing$,
    $\mathtt{UB}(\mathbf{x}) \gets \sum_{i=1}^{z} x_i \cdot \delta_{i}(e^{(i)})$\;
    \While{$\mathtt{UB}(\mathbf{x}) \geq \tau$}{
      $i^{*} \gets \argmax_{i \in [1,z]} x_i \cdot \delta_{i}(e^{(i)})$\;
      $\Delta_{e^{(i^{*})}} \gets \sum_{i=1}^{z} x_i \cdot \delta_i(e^{(i^{*})}),
      E_{\tau} \gets E_{\tau} \cup \{e^{(i^{*})}\}$\;
      $e^{(i^{*})} \gets \rl_{i^{*}}.\mathsf{next},
      \mathtt{UB}(\mathbf{x}) \gets \sum_{i=1}^{z} x_i \cdot \delta_{i}(e^{(i)})$\;
    }
    \Return{$E_{\tau}$}\;\label{line:mttd:retrieve:end}
  }
  \caption{\textsc{Multi-Topic ThresholdDescend}}
  \label{alg:mttd}
\end{algorithm}

The algorithmic description of \mttd is presented in Algorithm~\ref{alg:mttd}.
In the initialization phase (Lines~\ref{line:mttd:init:start}--\ref{line:mtts:init:end}),
the candidate $S$ and the element buffer $E'$ are both set to $\varnothing$.
The traversals of ranked lists are initialized in the same way as \mtts.
The initial threshold $\tau$ for the first round of evaluation is
the upper-bound score for any active element w.r.t.~$\mathbf{x}$
and the termination threshold $\tau'$ is $0$.
After initialization, \mttd runs each round of evaluation with threshold $\tau$
following Lines~\ref{line:mttd:round:start}--\ref{line:mttd:round:end}.
It first retrieves the set of elements $E_{\tau}$ whose scores
potentially reach $\tau$ from the ranked lists.
The method is shown in the procedure \texttt{retrieve($\tau$)}
(Lines~\ref{line:mttd:retrieve:start}--\ref{line:mttd:retrieve:end}),
which generally uses the same idea as \mtts:
it traverses each ranked list sequentially in decreasing order of topic-wise scores
and continuously adds the element with the maximum $x_i \cdot \delta_{i}(e^{(i)})$
to $E_{\tau}$ until the upper-bound score $\mathtt{UB}(\mathbf{x})$
is decreased to $\tau$.
After adding $E_{\tau}$ to the element buffer $E'$, the evaluation
procedure is started (Lines~\ref{line:mttd:eval:start}--\ref{line:mttd:eval:end}).
It always considers the element $e' \in E' \setminus S$ with the maximum $\Delta_{e'}$.
If the marginal gain $\Delta(e'|S)$ of adding $e'$ to $S$ is at least $\tau$,
$e'$ will be included into $S$ and deleted from $E'$.
When $S$ has contained $k$ elements, \mttd is directly terminated
and $S$ is returned as the result $S_{td}$ for $q_t(k,\mathbf{x})$.
The round of evaluation is finished when no elements in $E'$ could
achieve a marginal gain of $\tau$.
Next, the termination threshold $\tau'$ is updated and
the threshold $\tau$ is descended by $(1-\varepsilon)$ times
for the subsequent round of evaluation.
Finally, when $\tau$ is lower than $\tau'$,
no more rounds of evaluations are required.
In this case, $S$ is returned as the result $S_{td}$ for $q_t(k,\mathbf{x})$
even though it contains fewer than $k$ elements
(Line~\ref{line:mttd:result}).

\begin{figure}
  \centering
  \includegraphics[width=0.6\textwidth]{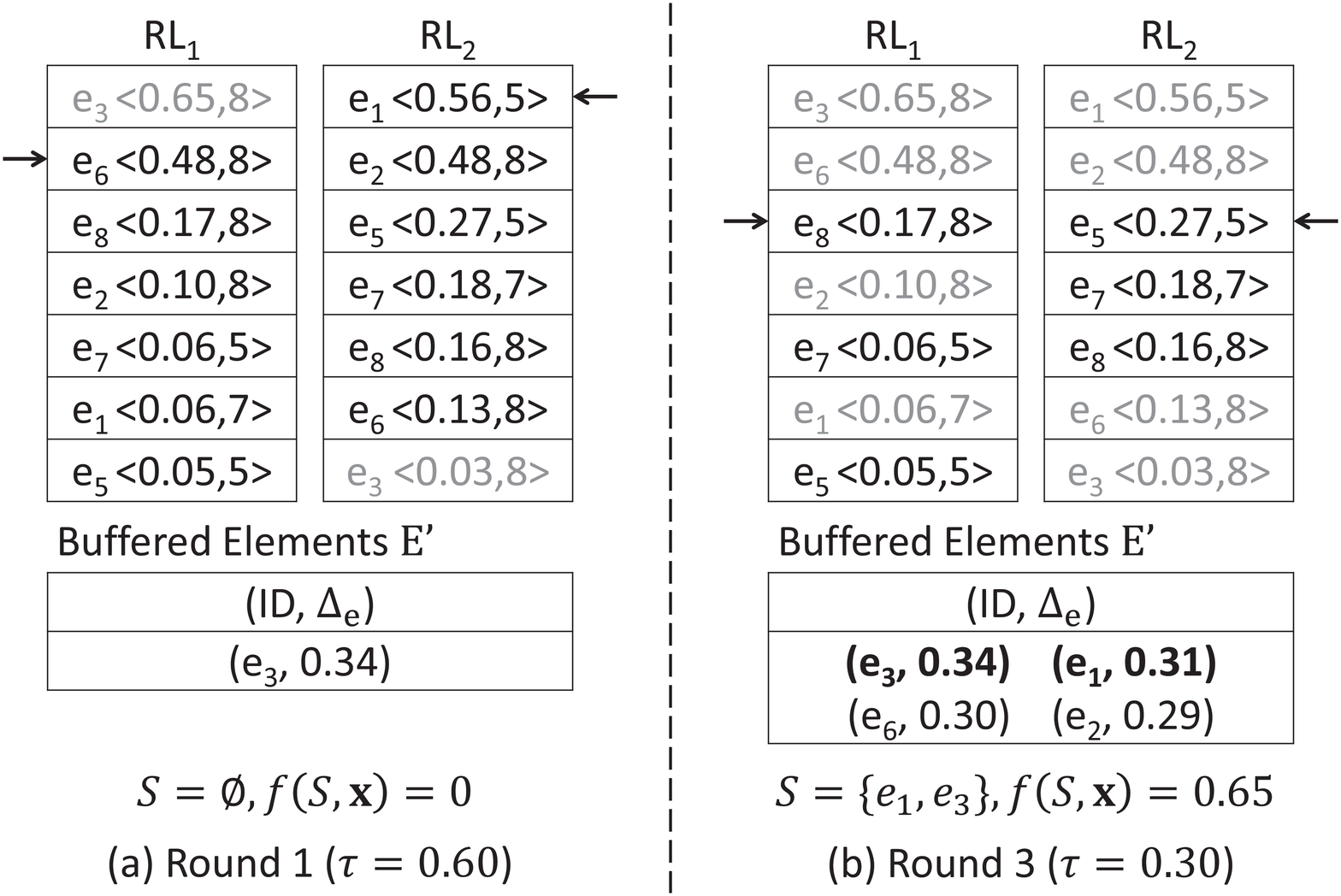}
  \caption{Example for \query processing using \mttd.}
  \label{fig:mttd}
\end{figure}

\begin{example}
In Figure~\ref{fig:mttd}, we illustrate the procedure for \mttd
to process a \query query $q_8(2,\mathbf{x})$
where $\mathbf{x}=(0.5,0.5)$
following the example in Table~\ref{tbl:example:definition}.
We also set $\varepsilon=0.3$ in this example.

First, \mttd initializes the threshold $\tau = 0.60$ for the first round
and the termination threshold $\tau'=0$.
The candidate $S$ and the element buffer $E'$ are initialized to $\varnothing$.
In Round 1 and 2 with $\tau=0.60$ and $0.42$, \mttd retrieves
$3$ elements $e_3, e_1, e_6$ from $\rl_{1}$ and $\rl_{2}$ and adds them to $E'$.
However, they are not evaluated in the first two rounds because
$\Delta_{e_3}=0.34$, $\Delta_{e_1}=0.31$, and $\Delta_{e_6}=0.30$,
all of which are smaller than $0.42$.
In Round 3 with $\tau=0.30$, $e_2$ is added to $E'$.
Then, $e_3$ is added to $S$ as $\Delta_{e_3}=0.34>\tau$.
Furthermore, $e_1$ is also added to $S$ as $\Delta_{e_1}=0.31>\tau$.
At this time, $S=\{e_1,e_3\}$ has contained two elements.
Therefore, \mttd is terminated and no more rounds are needed.
$S_{td}=\{e_1,e_3\}$ is returned as the result for $q_8(k,\mathbf{x})$.
\end{example}

The approximation ratio of \mttd is given in Theorem~\ref{thm:mttd:ratio}.

\begin{theorem}\label{thm:mttd:ratio}
The result $S_{td}$ returned by \mttd is $(1-\frac{1}{e}-\varepsilon)$-approximate for any \query query.
\end{theorem}
The proof is given in Appendix~\ref{subsec:proof:theorem:3}.

\textbf{Complexity Analysis.}
Let $\tau_0$ be the threshold $\tau$ of the first round in \mttd.
The number of rounds in \mttd is at most
$\lceil \log_{1-\varepsilon} (\frac{\tau'}{\tau_0}) \rceil$.
Because $\tau' = f(S,\mathbf{x}) \cdot \frac{\varepsilon}{k} \geq \delta_{max} \cdot \frac{\varepsilon}{k}$
and $\tau_0 \leq d \cdot \delta_{max}$,
we have $\frac{\tau_0}{\tau'} \leq \frac{kd}{\varepsilon}$
and the number of rounds is $\mathcal{O}(\frac{\log(kd)}{\varepsilon^{2}})$.
In each round, it evaluates $\mathcal{O}(n_t'')$ elements
where $n_t''$ is the number of elements in the buffer $E'$ of \mttd
and the evaluation of an element is also $\mathcal{O}(ld)$.
Here, we use a max-heap for $E'$ and thus it costs $\mathcal{O}(\log n_t'')$
to dequeue the top element from $E'$.
In addition, the time for retrieving an element from ranked lists is still $\mathcal{O}(\log n_t)$.
The complexity for each round is $\mathcal{O}\big(n_t'' \cdot (ld + \log n_t)\big)$.
Therefore, the time complexity of \mttd is
$\mathcal{O}\big( n_t'' \cdot \log(kd) \cdot \varepsilon^{-2} \cdot (ld + \log n_t)\big)$.

\section{Experiments}
\label{sec:experiments}

In this section, we conduct extensive experiments to verify
the effectiveness of \query query
as well as the efficiency of \mtts and \mttd for \query processing.
We first introduce the experimental setup in Section~\ref{subsec:exp:setup}.
Then, we show the results for the effectiveness of \query query in Section~\ref{subsec:exp:effectiveness}.
Finally, the results for the efficiency and scalability of \mtts and \mttd
are reported in Section~\ref{subsec:exp:efficiency}.

\subsection{Experimental Setup}
\label{subsec:exp:setup}

\begin{table}[t]
  \centering
  \small
  \caption{Statistics of datasets}
  \label{tbl:stat:datasets}
  \begin{tabular}{|c|c|c|c|}
    \hline
    \textbf{Dataset} & \textbf{AMiner} & \textbf{Reddit} & \textbf{Twitter} \\ \hline
    \textbf{Number of Elements} & 1.66M & 20.2M & 14.8M \\ \hline
    \textbf{Vocabulary Size} & 580K / 71K & 2.8M / 88K & 3.0M / 68K \\ \hline
    \textbf{Average Length} & 74.5 / 49.2 & 24.6 / 8.6 & 12.6 / 5.1 \\ \hline
    \textbf{Average References} & 3.68 & 0.85 & 0.62 \\ \hline
  \end{tabular}
\end{table}

\textbf{Dataset.}
Three real-world datasets used in the experiments are listed as follows.
\begin{itemize}
  \item \textbf{AMiner}~\cite{Tang:2009:SIA} is a collection of academic papers
  published in the ACM Digital Library till 2015. We assign random
  timestamps to the papers published in the same year.
  \item \textbf{Reddit}\footnote{https://www.reddit.com/r/datasets}
  is a collection of submissions and comments on Reddit from June 01, 2014 to June 14, 2014.
  \item \textbf{Twitter}\footnote{https://developer.twitter.com/en/docs}
  consists of the tweets collected via the streaming API from July 14, 2017 to July 26, 2017.
\end{itemize}
The statistics of the datasets are given in Table~\ref{tbl:stat:datasets}.
In the preprocessing, we remove stop words and noise words from the textual contents of elements.
Note that we report the vocabulary size and the average length of elements
both before and after the preprocessing.

\textbf{Topic Model.}
We use LDA~\cite{Blei:2003:LDA} to train topic models on the corpora of \emph{AMiner} and \emph{Reddit}.
PLDA~\cite{Liu:2011:PPL} is the implementation of LDA for training.
For topic training on the corpus of \emph{Twitter},
we use the biterm topic model~\cite{DBLP:conf/www/YanGLC13} (BTM)
because it is designed for short texts like tweets.
The corpus of each dataset consists of $e.doc$ of each element $e$.
To study how the number of topics $z$ affects the performance of compared methods,
we train 5 topic models for each dataset with $z$ ranging from $50$ to $250$.
Two Dirichlet priors $\alpha,\beta$ are set to $\frac{50}{z},0.01$
for both LDA and BTM.
The pre-trained topic models are loaded into memory and
used as a black-box oracle for each compared method.

\textbf{Compared Methods.}
We compare the following methods in Section~\ref{subsec:exp:effectiveness}
to evaluate the effectiveness of \query query.
\begin{itemize}
  \item \textbf{Top-$k$ Keyword Query} (TF-IDF) retrieves $k$ most relevant elements
  to the query keywords. We adopt the log-normalized TF-IDF weight
  to vectorize the elements and queries. Cosine similarity is used as the similarity
  measure between an element and a query.
  \item \textbf{Diversity-aware Top-$k$ Keyword Query}~\cite{Chen:2015:DTP} (DIV)
  considers both \textit{textual relevance} and \textit{result diversity}.
  Given a query $q$ and a set of elements $S$,
  we have $score(q,S) = \lambda \sum_{e \in S} rel(q,e) + (1-\lambda) div(S)$,
  where $rel(q,e)$ is the relevance of $e$ to $q$ and
  $div(S)$ is the average dissimilarity between each pair of elements in $S$.
  We set $\lambda=0.3$ following~\cite{Chen:2015:DTP}.
  A set of $k$ elements $S$ with the maximum $score(q,S)$ is returned as the result for $q$.
  \item \textbf{Sumblr}~\cite{DBLP:conf/sigir/ShouW0013} is a method for social stream summarization.
  In our experiments, we use Sumblr for query processing as follows:
  given a set of keywords, we select the elements that contain at least one keyword as candidates.
  Then, we run Sumblr on the candidates to generate a summary of $k$ elements as the query result.
  The parameters for k-means clustering and LexRank are the same as~\cite{DBLP:conf/sigir/ShouW0013}.
  \item \textbf{Top-$k$ Relevance Query}~\cite{Zhang:2017:PLQ} (REL) measures the relevance
  between an element and a query by topic modeling.
  It returns $k$ elements whose topic vectors have the highest
  cosine similarities to the query vector as the result.
  \item \textbf{\query Query} retrieves a set of
  elements $S$ maximizing $f(S,\mathbf{x})$
  w.r.t. a query vector $\mathbf{x}$. The results of
  \mttd are used in the effectiveness tests.
\end{itemize}
We note that TF-IDF, DIV, and Sumblr are keyword queries
while REL and \query use query vectors inferred from topic models.
To compare them fairly, the queries are generated as follows:
(1) draw the keywords from the vocabulary;
(2) acquire a query vector by treating the keywords as a pseudo-document
and inferring its topic vector from the topic model.
To retrieve the query results, TF-IDF, DIV, and Sumblr receive the keywords
while REL and \query receive the query vectors.

The following methods are compared in Section~\ref{subsec:exp:efficiency}
to evaluate their efficiency and scalability for \query processing.
\begin{itemize}
  \item \textbf{CELF}~\cite{Leskovec:2007:Cost} is an improved version
  of the basic greedy algorithm~\cite{Nemhauser:1978:Analysis}.
  It is the most common batch algorithm for submodular maximization
  and acquires $(1-\frac{1}{e})$-approximation results for \query queries.
  Note that $(1-\frac{1}{e})$ is the best approximation ratio
  for this problem unless P=NP~\cite{Feige:1998:TLN}.
  \item \textbf{SieveStreaming}~\cite{Badanidiyuru:2014:Streaming}
  is the state-of-the-art streaming algorithm for submodular maximization.
  It returns $(\frac{1}{2}-\varepsilon)$-approximation results for \query queries.
  \item \textbf{Top-$k$ Representative} retrieves
  $k$ elements with the highest representativeness scores $\delta(e,\mathbf{x})$
  w.r.t. a query vector $\mathbf{x}$ from ranked lists as the result,
  which is only $\frac{1}{k}$-approximate for \query queries.
  We compare with it to show that traditional methods for top-$k$ queries cannot work well for \query queries.
  \item \textbf{\mtts} and \textbf{\mttd} are our proposed algorithms for \query
  processing based on ranked lists.
\end{itemize}

\begin{table}
  \centering
  \small
  \caption{Parameters in the experiments}
  \label{tbl:parameters}
  \begin{tabular}{ccc}
    \hline
    \textbf{Parameter} & \textbf{Setting} & \textbf{Default} \\
    \hline
    the parameter $\varepsilon$ in \mtts/\mttd & 0.1 to 0.5 & 0.1 \\
    the result size $k$ & 5 to 25 & 10 \\
    the number of topics $z$ & 50 to 250 & 50 \\
    the window length $T$ & 6 hours to 30 hours & 24 hours \\
    \hline
  \end{tabular}
\end{table}

\textbf{Query and Workload Generation.}
We generate a \query query as follows:
(1) draw 1--5 words randomly from the vocabulary;
(2) acquire the query vector by inferring
the topic distribution of selected words from the
topic model.

In an experiment, we feed all elements in a dataset to
compared methods in ascending order of timestamp.
The active window and ranked lists perform batch-updates
for each bucket of elements.
Then, the query workload is generated as follows:
we generate 10K \query queries for each dataset
and assign a random timestamp in range $[1,t_n]$
($t_n$ is the end time of the stream) to each query.
The query results are retrieved at the assigned timestamps.

\textbf{Parameter Setting.}
The parameters we examine in the experiments
are listed in Table~\ref{tbl:parameters}.
In addition, the factors $\lambda,\eta$ in Equation~\ref{eq:topic-utility}
are set to $0.5,20$ for the \textit{AMiner} and \textit{Reddit} datasets,
and $0.5,200$ for the \textit{Twitter} dataset.
The bucket length $L$ is fixed to $15$ minutes.

\textbf{Experimental Environment.}
All experiments are conducted on a server running Ubuntu 16.04.3 LTS.
It has an Intel Xeon E7-4820 1.9GHz processor and 128 GB memory.
All compared methods are implemented in Java 8.

\subsection{Effectiveness}
\label{subsec:exp:effectiveness}

\begin{table}[t]
  \centering
  \small
  \caption{Results for user study}
  \label{tbl:userstudy}
  \begin{tabular}{|c|c|c|c|c|c|c|}
    \hline
    \multicolumn{2}{|c|}{\textbf{Method}} & TF-IDF & DIV & Sumblr & REL & \query \\ \hline
    \multirow{2}{*}{\textbf{AMiner}}
    & Represent. & 2.28 & 1.56 & 3.72 & 2.78 & \textbf{4.67} \\ \cline{2-7}
    & Impact     & 2.39 & 1.44 & 4.01 & 2.39 & \textbf{4.78} \\ \hline
    \multirow{2}{*}{\textbf{Reddit}}
    & Represent. & 2.05 & 3.00 & 3.67 & 1.95 & \textbf{4.33} \\ \cline{2-7}
    & Impact     & 1.80 & 2.24 & 3.80 & 2.33 & \textbf{4.80} \\ \hline
    \multirow{2}{*}{\textbf{Twitter}}
    & Represent. & 1.79 & 2.38 & 4.08 & 2.08 & \textbf{4.67} \\ \cline{2-7}
    & Impact     & 1.58 & 2.25 & 4.01 & 2.34 & \textbf{4.88} \\ \hline
  \end{tabular}
\end{table}

To evaluate the effectiveness of our \query query, we first conduct
a study on users' satisfaction for the results returned by
each query method.
We follow the methodology and procedure of user study in previous work
on social search~\cite{Chen:2015:DTP}. The detailed procedure
is as follows.

First, we generate 20 queries by selecting 20 trending topics on three datasets
(e.g., ``\emph{social media analysis}'' on \emph{AMiner}, ``\emph{NBA}'' on \emph{Reddit}, and
``\emph{pop music}'' on \emph{Twitter}) and use the topical words of each topic as keywords.
Second, we process these queries with each method in the default setting
and return a set of five elements as the results.
Third, we recruit 30 volunteers who are not related to this work
and familiar with the query topics
to evaluate the result quality of compared methods.
For each query, we ask 3 different evaluators to rank the quality of result sets
and record the average score on each aspect.
Specifically, each evaluator is requested to rank his/her satisfaction for
the result sets on two aspects:
(1) \emph{representativeness}:
the relevance to query topic and the information coverage on the query topic of its entirety
(ranking from ``the least representative'' to ``the most representative'', mapped to values $1$ to $5$);
(2) \emph{impact}: the number of citations, comments, and retweets of selected elements
(ranking from ``the lowest impact'' to ``the highest impact'', mapped to values $1$ to $5$).

The results of the user study are shown in Table~\ref{tbl:userstudy}.
Following~\cite{Chen:2015:DTP}, we measure the agreement between different
users by computing the Cohen's linearly weighted kappa~\cite{Cohen:1968:Weighted}
for each query on each aspect.
The kappa values for \emph{representativeness} are between $0.5$ and $0.89$ ($0.72$ on average).
The kappa values for \emph{impact} are in the range
of $0.56$--$1.0$ ($0.79$ on average).
We observe that \query achieves the highest scores among compared methods
on both \emph{representativeness} and \emph{impact} in all datasets.
We also collect feedback from users for the reason of dissatisfaction.
``Low coverage'' is the primary problem for TF-IDF and REL,
while ``containing irrelevant elements'' is the main reason
why the results of DIV and Sumblr are unsatisfactory.

\begin{table}[t]
  \centering
  \small
  \caption{Results for quantitative analysis}
  \label{tbl:effectiveness}
  \begin{tabular}{|c|c|c|c|c|c|c|}
    \hline
    \multicolumn{2}{|c|}{\textbf{Method}} & TF-IDF & DIV & Sumblr & REL & \query \\ \hline
    \multirow{2}{*}{\textbf{AMiner}}
    & Coverage  & 0.1968 & 0.1766 & 0.2140 & 0.2400 & \textbf{0.2663} \\ \cline{2-7}
    & Influence & 0.0765 & 0.0777 & 0.5470 & 0.1159 & \textbf{0.8430} \\ \hline
    \multirow{2}{*}{\textbf{Reddit}}
    & Coverage  & 0.2387 & 0.2050 & 0.2419 & 0.2885 & \textbf{0.3162} \\ \cline{2-7}
    & Influence & 0.0175 & 0.0107 & 0.4315 & 0.0143 & \textbf{0.5862} \\ \hline
    \multirow{2}{*}{\textbf{Twitter}}
    & Coverage  & 0.2200 & 0.2118 & 0.2213 & 0.2722 & \textbf{0.3052} \\ \cline{2-7}
    & Influence & 0.0295 & 0.0296 & 0.1611 & 0.1268 & \textbf{0.6516} \\ \hline
  \end{tabular}
\end{table}

Then, we use two quantitative metrics to
evaluate the effectiveness of \query query:
(1) \emph{coverage}: do the result sets achieve high information coverage on query topics?
Following the metric used in previous studies~\cite{Lin:2010:Multi,Badanidiyuru:2014:Streaming},
the coverage score of a result set $S$ w.r.t. a query vector $\mathbf{x}$ is computed by
$\sum_{e \in A_t \setminus S} \max_{e' \in S}$ $rel(e,\mathbf{x}) \cdot sim(e,e')$
where $rel(e,\mathbf{x})$ is the relevance of $e$ to $\mathbf{x}$ and
$sim(e,e')$ is the similarity of $e$ and $e'$;
(2) \emph{influence}: are the result sets referred by a large number of elements
(e.g., citations, comments, retweets, and so on)?
We use the total number of elements referring to at least one element
in the result set as the influence score.
For ease of presentation, the influence scores are
linearly scaled to $[0,1]$ by dividing by
the influence score of top-$k$ influential elements.
To acquire the results shown in Table~\ref{tbl:effectiveness},
we sample the result sets of 1K queries returned by each method
and compute the average scores.

We present the quantitative results for the effectiveness of
compared methods in Table~\ref{tbl:effectiveness}.
First, \query outperforms other query methods on \textit{information coverage},
which verifies that our semantic model is able to preserve information on query topics.
Second, as only \query and Sumblr account for the influences of elements,
they naturally achieve much higher influence scores than other methods.
\query further outperforms Sumblr in terms of \textit{influence}
because \query directly adopt the number of references for influence computation
while Sumblr only considers the PageRank scores of authors.

Overall, the above results have confirmed that
\query shows better result quality than existing methods for social search and summarization
in terms of \emph{information coverage} and \emph{influence}.

\begin{figure}
  \centering
  \begin{minipage}{0.64\textwidth}
    \centering
    \includegraphics[width=\textwidth]{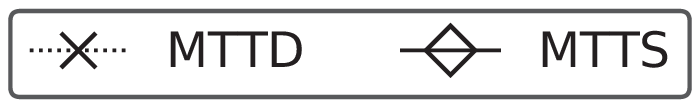}
  \end{minipage}%
  \\
  \begin{minipage}{0.75\textwidth}
    \centering
    \includegraphics[width=\textwidth]{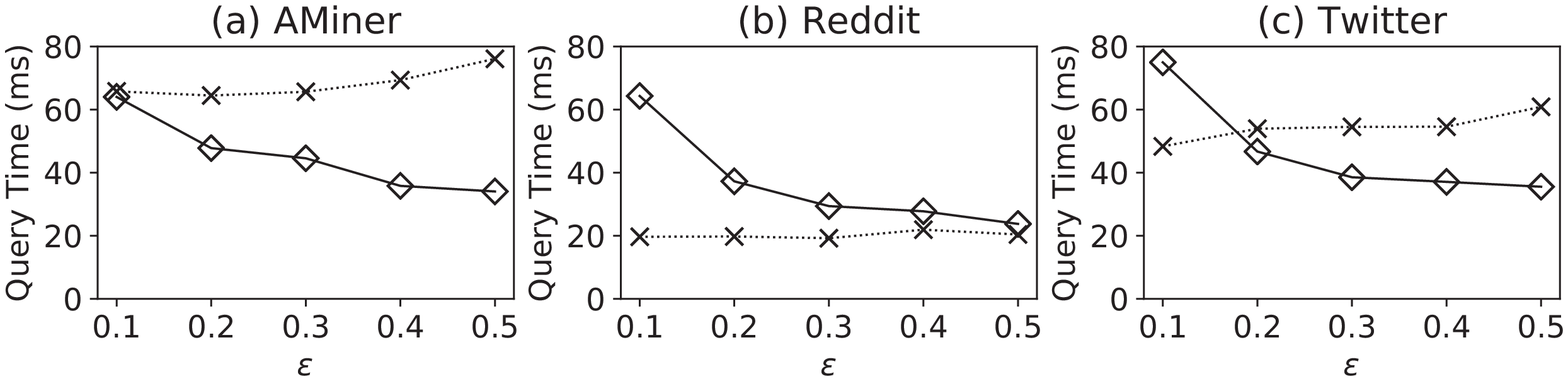}
    \vspace{-2em}
    \caption{Query time with varying $\varepsilon$}
    \label{fig:time:epsilon}
  \end{minipage}%
  \\ \vspace{1em}
  \begin{minipage}{0.75\textwidth}
    \centering
    \includegraphics[width=\textwidth]{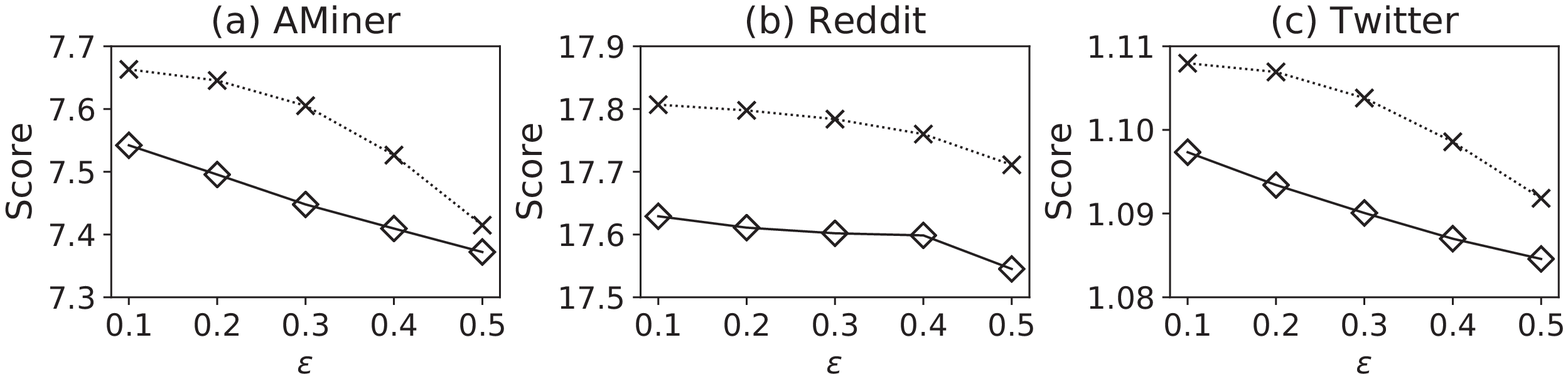}
    \vspace{-2em}
    \caption{Scores with varying $\varepsilon$}
    \label{fig:score:epsilon}
  \end{minipage}%
\end{figure}

\subsection{Efficiency and Scalability}
\label{subsec:exp:efficiency}

\textbf{Effect of $\varepsilon$.}
The average CPU time of \mtts and \mttd to process one \query query
(i.e., \textit{query time}) with varying $\varepsilon$
is illustrated in Figure~\ref{fig:time:epsilon}.
\mtts and \mttd show different trends w.r.t.~$\varepsilon$.
On the one hand, the query time of \mtts drops drastically
when $\varepsilon$ increases as the number of candidates
in \mtts is inversely proportional to $\varepsilon$.
On the other hand, \mttd is not sensitive to $\varepsilon$ and
typically takes slightly more time for a larger $\varepsilon$.
This is because a greater $\varepsilon$ often leads to a smaller
threshold for termination. In this case, more elements are retrieved
from ranked lists and evaluated by \mttd, which degrades the query efficiency.

The average scores of the results returned by \mtts and \mttd
with varying $\varepsilon$ are shown in Figure~\ref{fig:score:epsilon}.
The scores of both methods decrease when $\varepsilon$ increases,
which is consistent with the theoretical results of
Theorem~\ref{thm:mtts:ratio} and~\ref{thm:mttd:ratio}.
However, both methods show good robustness against $\varepsilon$:
compared with CELF, their quality losses are at most 5\%
even when $\varepsilon=0.5$.

\begin{figure}
  \centering
  \begin{minipage}{0.64\textwidth}
    \centering
    \includegraphics[width=\textwidth]{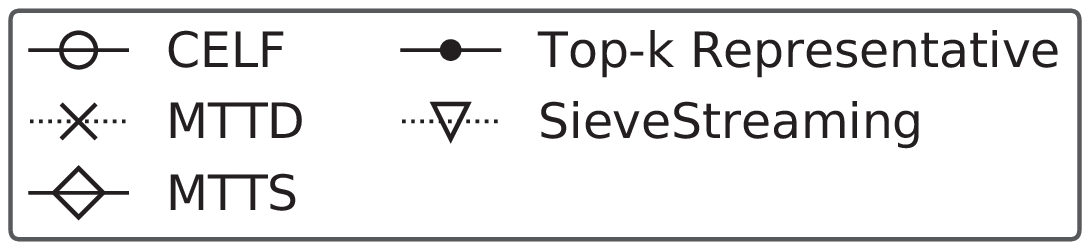}
  \end{minipage}%
  \\
  \begin{minipage}{0.75\textwidth}
    \centering
    \includegraphics[width=\textwidth]{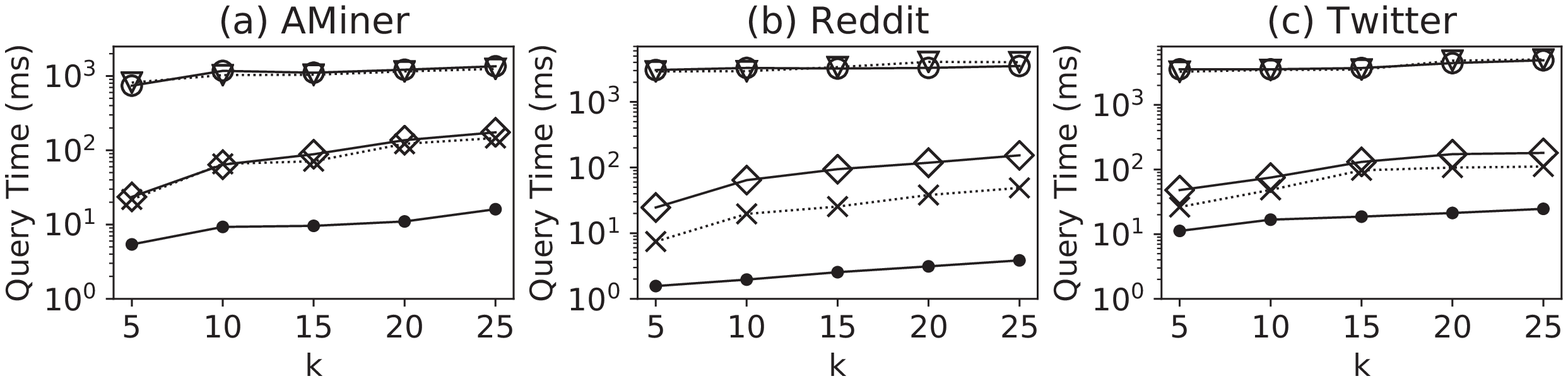}
    \vspace{-2em}
    \caption{Query time with varying $k$}
    \label{fig:time:k}
  \end{minipage}%
  \\ \vspace{1em}
  \begin{minipage}{0.75\textwidth}
    \centering
    \includegraphics[width=\textwidth]{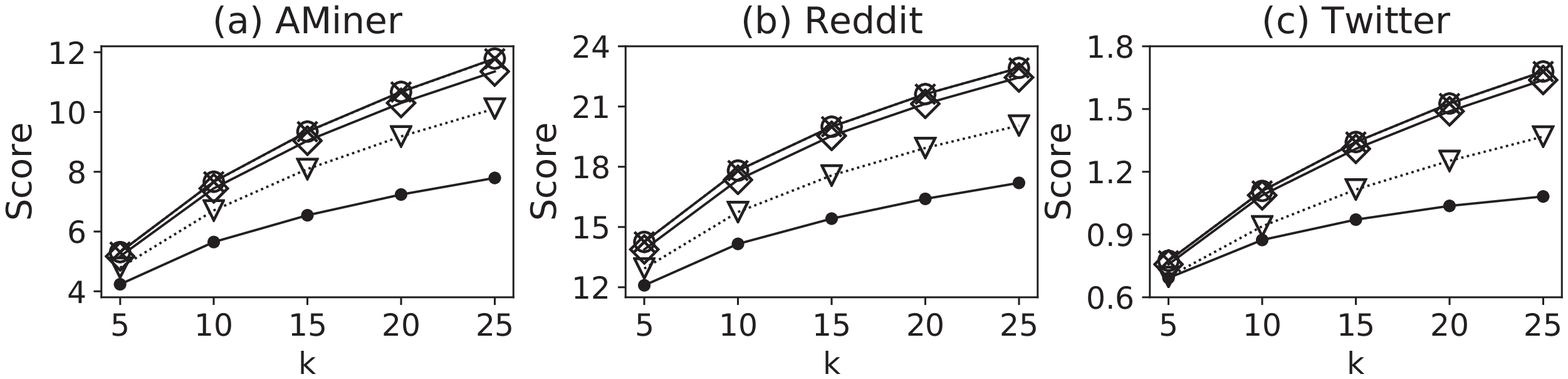}
    \vspace{-2em}
    \caption{Scores with varying $k$}
    \label{fig:score:k}
  \end{minipage}%
  \\ \vspace{1em}
  \begin{minipage}{0.75\textwidth}
    \centering
    \includegraphics[width=\textwidth]{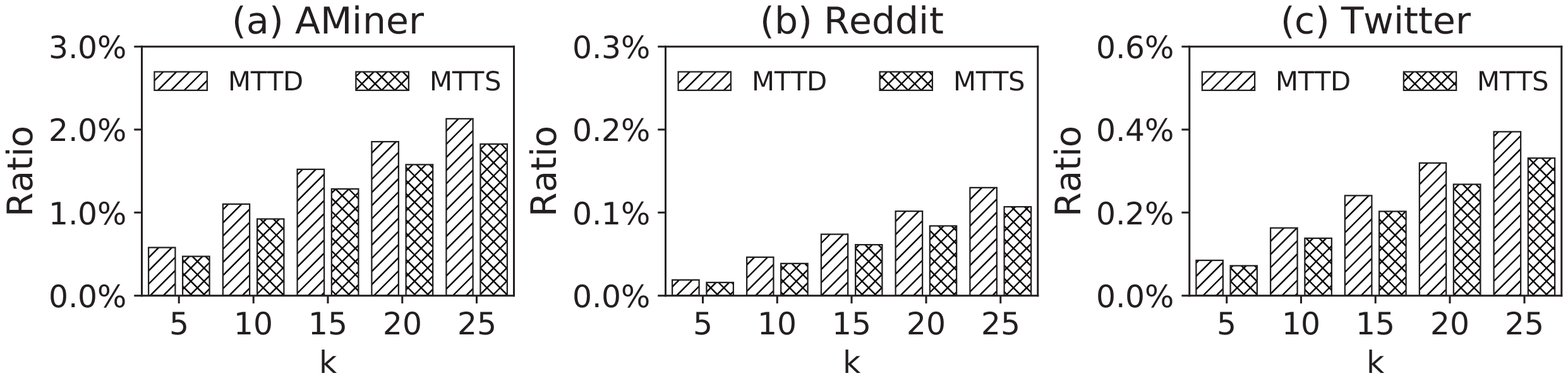}
    \vspace{-2em}
    \caption{Ratios of evaluated elements with varying $k$}
    \label{fig:ratio:k}
  \end{minipage}%
\end{figure}

\textbf{Effect of result size $k$.}
The average query time of compared methods with varying $k$
is presented in Figure~\ref{fig:time:k}.
In addition, the average ratios between the number of elements
evaluated by \mtts/\mttd and the number of active elements
are shown in Figure~\ref{fig:ratio:k}.
First of all, \mtts and \mttd run at least one order of magnitude
faster than CELF and SieveStreaming
for \query processing in all datasets.
\mtts and \mttd can achieve up to $124$x and $390$x speedups over
the two baselines respectively. Compared with them,
\mtts and \mttd can prune most of the unnecessary evaluations
(at least $98\%$ as shown in Figure~\ref{fig:ratio:k})
by utilizing the ranked lists.
Then, the query time of \mtts and \mttd significantly grows with increasing $k$.
The result can be explained by the ratios of evaluated elements.
From Figure~\ref{fig:ratio:k}, we can see the ratio increases near linearly with $k$.
As more elements are evaluated when $k$ increases,
the query time naturally rises.
Finally, we can see \mttd outperforms \mtts in most cases but
the ratio of elements evaluated by \mttd is always higher than \mtts.
This is because \mttd only keeps one candidate but \mtts maintains
multiple candidates independently. As a result, \mttd reduces the number
of evaluations though it retrieves more elements from ranked lists than \mtts.

The average scores of the results returned by \mtts and \mttd
with varying $k$ are shown in Figure~\ref{fig:score:k}.
We can see the result quality of \mttd is always nearly equal (>$99\%$)
to CELF for different $k$. Meanwhile, \mtts can also return results
with over $95\%$ representativeness scores compared with CELF.
The results of SieveStreaming are inferior to those of CELF, \mtts, and \mttd.
Although Top-$k$ Representative shows the best performance among compared methods,
its results are of the lowest quality among compared methods.
In addition, its result quality degrades dramatically when $k$ increases
because the word and influence overlaps are ignored.

\begin{figure}
  \centering
  \begin{minipage}{0.64\textwidth}
    \centering
    \includegraphics[width=\textwidth]{legend-k-z-T}
  \end{minipage}%
  \\
  \begin{minipage}{0.75\textwidth}
    \centering
    \includegraphics[width=\textwidth]{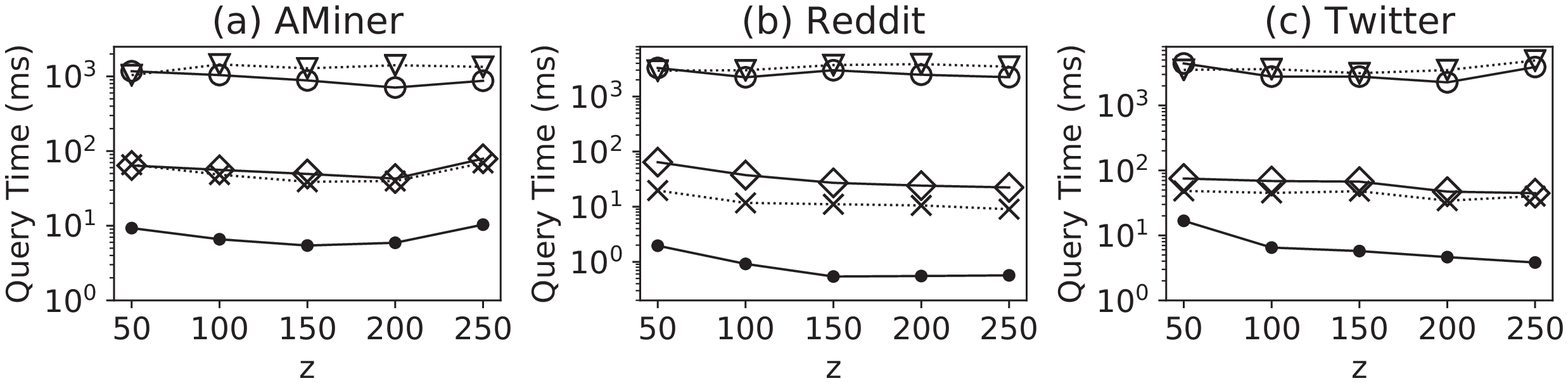}
    \vspace{-2em}
    \caption{Query time with varying $z$}
    \label{fig:time:z}
  \end{minipage}%
  \\ \vspace{1em}
  \begin{minipage}{0.75\textwidth}
    \centering
    \includegraphics[width=\textwidth]{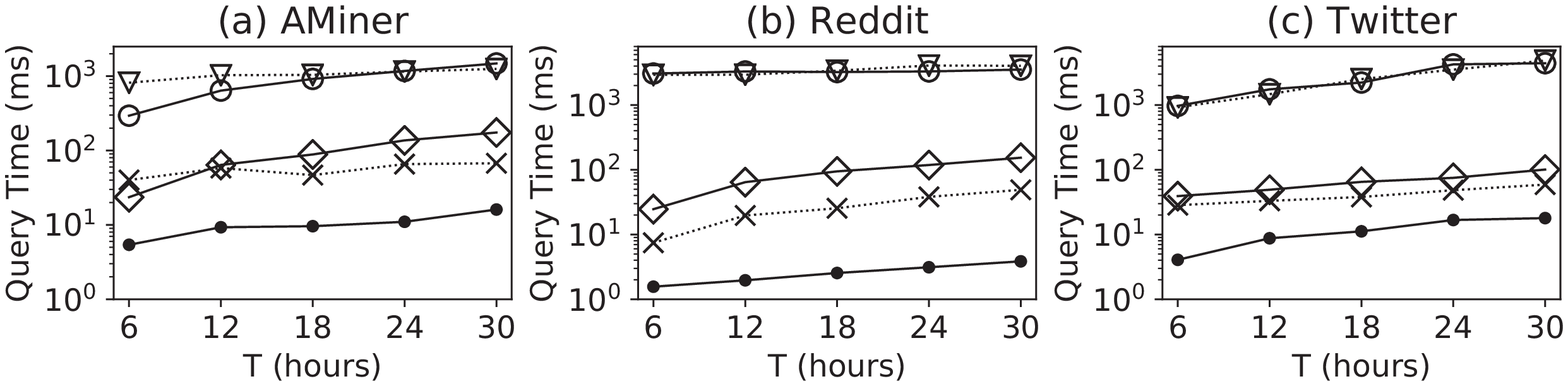}
    \vspace{-2em}
    \caption{Query time with varying $T$}
    \label{fig:time:T}
  \end{minipage}%
  \\ \vspace{1em}
  \begin{minipage}{0.75\textwidth}
    \centering
    \includegraphics[width=\textwidth]{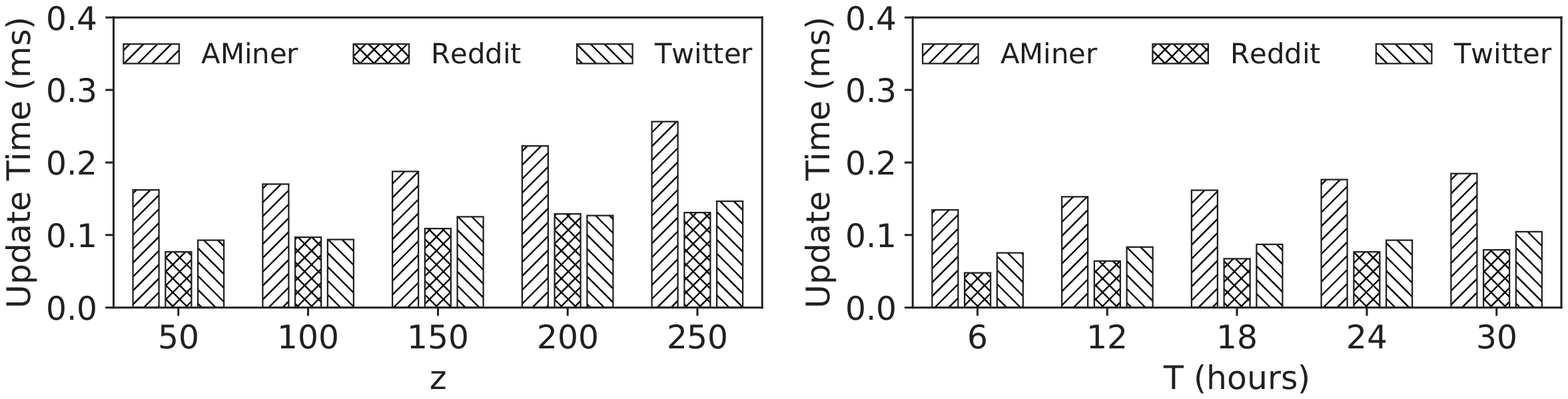}
    \vspace{-2em}
    \caption{Update time with varying $z$ and $T$}
    \label{fig:update:time}
  \end{minipage}%
\end{figure}

\textbf{Scalability.}
We evaluate the scalability of \mtts and \mttd with varying the number of topics $z$
and the window length $T$. The results for query time are illustrated
in Figure~\ref{fig:time:z} and~\ref{fig:time:T}.
The query time of \mtts and \mttd drops when $z$ increases.
Because the average number of elements on each topic deceases with increasing $z$,
the number of evaluated elements naturally decreases.
However, when $z=250$ in the \emph{AMiner} dataset, the query time of \mtts and \mttd
grows because there are more non-zero entries in the query vectors.
The query time of all methods increases with $T$
since there are more active elements.
Nevertheless, \mtts and \mttd significantly outperform the baselines in all cases.

The average CPU time elapsed to update the ranked lists per arrival element
is shown in Figure~\ref{fig:update:time}.
We can see it takes more update time when $z$ or $T$ increases.
As the number of maintained ranked lists is equal to $z$ and
the number of active elements grows with $T$,
the cost for ranked list maintenance inevitably rises with increasing $z$ or $T$.
Nevertheless, the update time is always lower than $0.3$ms in all datasets.

Overall, the experimental results show that our proposed methods
demonstrate high efficiency and scalability for both ranked list maintenance
and \query processing, which can meet the requirements for real-world
social streams.

\section{Conclusion}

In this paper, we defined a novel \query query to retrieve
a set of $k$ representative elements from a social stream w.r.t.~a query vector.
We then proposed two algorithms, namely \mtts and \mttd, that leveraged the ranked lists
for \query processing over sliding windows. Theoretically, \mtts and \mttd provided
$(\frac{1}{2}-\varepsilon)$ and $(1-\frac{1}{e}-\varepsilon)$ approximation results
for \query queries respectively. Finally, we conducted extensive experiments on real-world
datasets to demonstrate that (1) the \query query achieved better performance in terms of
\emph{information coverage} and \emph{social influence}
than existing query methods on social data;
(2) \mtts and \mttd had much higher efficiency and scalability than
the baselines for \query processing with near-equivalent result quality.
In future work, we plan to extend our approach for
supporting the incremental updates of topic models over streams.

\bibliographystyle{plainnat}
\bibliography{references}

\appendix
\section{Proofs of Lemmas and Theorems}

\subsection{Proof of Lemma~\ref{prop:semantic:submodular}}
\label{subsec:proof:lemma:1}

\begin{proof}
First of all, for $e \in E \setminus S$ and $S \subseteq E$,
we have
\begin{equation*}
  \mathcal{R}_{i}(S\cup\{e\})-\mathcal{R}_{i}(S) \geq \sum_{w \in V_e \setminus V_S} \sigma_{i}(w,e) \geq 0
\end{equation*}
because $-p \cdot \log p \geq 0$ for $p \in [0,1]$.
Thus, $\mathcal{R}_{i}(\cdot)$ is monotone.

Given any $e \in E \setminus T$ and $S \subseteq T \subseteq E$,
we use $\Delta(e|S) = \mathcal{R}_{i}(S\cup\{e\}) - \mathcal{R}_{i}(S)$
and $\Delta(e|T) = \mathcal{R}_{i}(T\cup\{e\}) - \mathcal{R}_{i}(T)$
to denote the marginal score gains of adding $e$ to $S$ and $T$.
Firstly, as $S \subseteq T$, $V_S \subseteq V_T$.
We divide $V_e$ into three disjoint subsets:
\begin{equation*}
  V_1 = V_e \setminus V_T,\quad V_2=V_e \cap (V_T \setminus V_S),\quad V_3=V_e \cap V_S
\end{equation*}
Then, it is obvious that
\begin{equation*}
  \Delta(e|\cdot)=\Delta(V_1|\cdot)+\Delta(V_2|\cdot)+\Delta(V_3|\cdot)
\end{equation*}
for $S$ and $T$.
For $V_1$, we have
\begin{equation*}
  \Delta(V_1|S)=\Delta(V_1|T)=\sum_{w\in V_1}\sigma_{i}(w,e)
\end{equation*}
because $V_1 \cap V_S = \varnothing$ and $V_1 \cap V_T = \varnothing$.
For $V_2$, we have
\begin{equation*}
  \Delta(V_2|S)=\sum_{w \in V_2} \sigma_{i}(w,e), \quad
  \Delta(V_2|T)=\sum_{w \in V_2} \max\big(0,\sigma_{i}(w,e)-\max_{e' \in T}\sigma_{i}(w,e')\big)
\end{equation*}
as $V_2 \cap V_S = \varnothing$ and $V_2 \subseteq V_T$.
Obviously, we can acquire $\Delta(V_2|S) \geq \Delta(V_2|T)$ as well.
For $V_3$, we have
\begin{equation*}
  \Delta(V_3|S) = \sum_{w \in V_3} \max \big(0,\sigma_{i}(w,e) - \max_{e' \in S} \sigma_{i}(w,e')\big), \quad
  \Delta(V_3|T) = \sum_{w \in V_3} \max\big(0,\sigma_{i}(w,e) -\max_{e' \in T}\sigma_{i}(w,e')\big)
\end{equation*}
because of $V_3 \subseteq V_S \subseteq V_T$.
Because $\max_{e' \in S}$ $\sigma_{i}(w,e') \leq \max_{e' \in T} \sigma_{i}(w,e')$
for $S \subseteq T$, $\Delta(V_3|S) \geq \Delta(V_3|T)$.
According to the above results, we prove $\Delta(e|S)\geq\Delta(e|T)$
and thus $\mathcal{R}_{i}(\cdot)$ is submodular.
\end{proof}

\subsection{Proof of Lemma~\ref{prop:influence:submodular}}
\label{subsec:proof:lemma:2}

\begin{proof}
First, given any $e'\in E \setminus S$ and $S \subseteq E$, for each $e \in I_t(S)$,
we have
\begin{equation*}
\begin{split}
  p_{i}(S \cup \{e'\} \leadsto e) - p_{i}(S \leadsto e)
  & = 1-\big(1-p_{i}(S \leadsto e)\big)\cdot\big(1-p_{i}(e' \leadsto e)\big) - p_{i}(S \leadsto e) \\
  & = p_{i}(e' \leadsto e)\cdot\big(1-p_{i}(S \leadsto e)\big)\geq 0
\end{split}
\end{equation*}
for $p_{i}(S \leadsto e) \in [0,1]$.

Second, given any $S \subseteq T \subseteq E$, for each $e \in I_t(T)$,
we have $p_{i}(S \leadsto e) \leq p_{i}(T \leadsto e)$ for $e.ref \cap I_t(S) \subseteq e.ref \cap I_t(T)$.
Therefore, for any $e'\in E \setminus T$,
we have
\begin{equation*}
\begin{split}
  p_{i}(S\cup\{e'\} \leadsto e)-p_{i}(S \leadsto e)
  & = 1-\big(1-p_{i}(S \leadsto e)\big)\cdot\big(1-p_{i}(e' \leadsto e)\big)-p_{i}(S \leadsto e) \\
  & = p_{i}(e' \leadsto e)\cdot\big(1-p_{i}(S \leadsto e)\big) \\
  & \geq p_{i}(e' \leadsto e)\cdot\big(1-p_{i}(T \leadsto e)\big) \\
  & = p_{i}(T\cup\{e'\} \leadsto e)-p_{i}(T \leadsto e)
\end{split}
\end{equation*}
Finally, because $\mathcal{I}_{i,t}(S)=\sum_{e \in I_t(S)} p_{i}(S \leadsto e)$ and
$p_{i}(\cdot \leadsto e)$ is monotone and submodular,
$\mathcal{I}_{i,t}(\cdot)$ is monotone and submodular as well.
\end{proof}

\subsection{Proof of Theorem~\ref{thm:mtts:ratio}}
\label{subsec:proof:theorem:1}

\begin{proof}
The sequence of estimations $\Phi$ for $\mathtt{OPT}$ is
in range $[\delta_{max},2\cdot k \cdot \delta_{max}]$.
Due to the monotonicity and submodularity of $f(\cdot,\mathbf{x})$,
we have $\mathtt{OPT} \in [\delta_{max}, k \cdot \delta_{max}]$.
Therefore, there must exist some $\phi\in\Phi$ such that
$(1-\varepsilon)\mathtt{OPT} \leq \phi \leq \mathtt{OPT}$.

Next, we discuss two cases for such $\phi$ and $S_{\phi}$.

\noindent\textbf{Case 1 ($|S_{\phi}|=k$).}
For each $e \in S_{\phi}$, we have $\Delta(e|S') \geq \frac{\phi}{2k}$ where
$S'$ is the subset of $S_{\phi}$ when $e$ is added.
Therefore,
\begin{equation*}
  f(S_{\phi},\mathbf{x}) \geq k \cdot \frac{\phi}{2k} \geq (\frac{1}{2}-\varepsilon) \mathtt{OPT}
\end{equation*}

\noindent\textbf{Case 2 ($|S_{\phi}|<k$).}
For each $e \in S^{*} \setminus S_{\phi}$, if $e$ has been evaluated by \mtts,
it is excluded from $S_{\phi}$ because $\Delta(e|S')<\frac{\phi}{2k}$
where $S'$ is the subset of $S_{\phi}$ when $e$ is evaluated;
if $e$ has not been evaluated by \mtts, it holds that
$\Delta(e|S)\leq\delta(e,\mathbf{x})<\mathtt{UB}(\mathbf{x})<\mathtt{TH}\leq\frac{\phi}{2k}$.
Thus,
\begin{equation*}
\begin{split}
  \mathtt{OPT} - f(S_{\phi},\mathbf{x})
  & \leq f(S^{*} \cup S_{\phi},\mathbf{x}) - f(S_{\phi},\mathbf{x}) \\
  & \leq \sum_{e \in S^{*} \setminus S_{\phi}} \Delta(e|S_{\phi}) \\
  & \leq k \cdot \frac{\phi}{2k} \leq \frac{1}{2} \cdot \mathtt{OPT}
\end{split}
\end{equation*}
Equivalently, $f(S_{\phi},\mathbf{x}) \geq \frac{1}{2} \cdot \mathtt{OPT}$.

In both cases, we have
$f(S_{ts},\mathbf{x}) \geq f(S_{\phi},\mathbf{x}) \geq (\frac{1}{2}-\varepsilon) \mathtt{OPT}$.
\end{proof}

\subsection{Proof of Theorem~\ref{thm:mttd:ratio}}
\label{subsec:proof:theorem:3}

\begin{proof}
There are two cases when \mttd is terminated.
Here, we discuss them separately.

\noindent\textbf{Case 1 $(|S_{td}|=k)$.}
Let $S_j=\{e_1,\ldots,e_j\}$ ($j\in[1,k]$) be the subset of $S_{td}$
after the first $j$ elements are added and $S_0 = \varnothing$.
Assume that $e_{j+1}$ is added to $S_j$ in the round with threshold $\tau$.
It holds that
\begin{equation*}
  \Delta(e_{j+1}|S_j) \geq \tau, \quad \Delta(e|S_j) < \frac{\tau}{1-\varepsilon}, \forall e \notin S_j \cup \{e_{j+1}\}
\end{equation*}
Then, we have
\begin{equation*}
  \Delta(e_{j+1}|S_j) \geq (1-\varepsilon) \Delta(e|S_j), \forall e \in S^*\setminus S_j
\end{equation*}
By summing up the above inequality for $e \in S^*\setminus S_j$,
we have
\begin{equation*}
  |S^*\setminus S_j| \cdot \Delta(e_{j+1}|S_j) \geq (1-\varepsilon) \sum_{e \in S^*\setminus S_j} \Delta(e|S_j)
\end{equation*}
Thus, we get
\begin{equation*}
  \Delta(e_{j+1}|S_j) \geq \frac{1-\varepsilon}{|S^*\setminus S_j|} \cdot \sum_{e \in S^*\setminus S_j} \Delta(e|S_j)
  \geq \frac{1-\varepsilon}{k} \cdot \sum_{e \in S^*\setminus S_j} \Delta(e|S_j)
\end{equation*}
Due to the submodularity of $f(\cdot,\mathbf{x})$, we have
$\sum_{e \in S^*\setminus S_j} \Delta(e|S_j)  \geq \mathtt{OPT} - f(S_j,\mathbf{x})$.
Thus,
\begin{equation*}
  \Delta(e_{j+1}|S_j) = f(S_{j+1},\mathbf{x}) - f(S_{j},\mathbf{x}) \geq \frac{1-\varepsilon}{k} (\mathtt{OPT} - f(S_j,\mathbf{x}))
\end{equation*}
Equivalently, we acquire
\begin{equation*}
  f(S_{j+1},\mathbf{x}) - \mathtt{OPT} \geq (1-\frac{1-\varepsilon}{k}) (f(S_{j},\mathbf{x}) - \mathtt{OPT})
\end{equation*}
Substituting $S_{j+1}$ by $S_{k},\ldots,S_1$
for $k$ times, we prove
\begin{equation*}
  f(S_{td},\mathbf{x}) = f(S_k,\mathbf{x})
  \geq \big( 1-(1-\frac{1-\varepsilon}{k})^{k} \big) \cdot \mathtt{OPT}
  \geq (1-e^{-(1-\varepsilon)})\mathtt{OPT}
  \geq (1-\frac{1}{e}-\varepsilon)\mathtt{OPT}
\end{equation*}

\noindent\textbf{Case 2 $(|S_{td}|<k)$.}
We have
\begin{equation*}
  \Delta(e|S_{td}) < \tau' = f(S_{td},\mathbf{x}) \cdot \frac{\varepsilon}{k}, \forall e \in S^*\setminus S_{td}
\end{equation*}
Therefore,
\begin{equation*}
  \mathtt{OPT}-f(S_{td},\mathbf{x}) \leq \sum_{e \in S^*\setminus S_{td}} \Delta(e|S_{td})
  \leq \sum_{e \in S^*\setminus S_{td}} f(S_{td},\mathbf{x}) \cdot \frac{\varepsilon}{k}
  \leq \varepsilon \cdot f(S_{td},\mathbf{x})
\end{equation*}
Thus, we acquire
\begin{equation*}
  f(S_{td},\mathbf{x}) \geq \frac{\mathtt{OPT}}{1+\varepsilon} \geq (1-\varepsilon)\mathtt{OPT}
\end{equation*}

In both cases, $f(S_{td},\mathbf{x}) \geq (1-\frac{1}{e}-\varepsilon)\mathtt{OPT}$.
\end{proof} 

\end{document}